\documentclass[%
preprint,superscriptaddress, amsmath,amssymb, aps,]{revtex4-1}

\usepackage{graphicx}
\usepackage{dcolumn}
\usepackage{bm}
\usepackage{slashed}
\usepackage{subfigure}
\usepackage[colorlinks=true,linkcolor=red,citecolor=blue]{hyperref}

\begin{document}

\title{Three-body decays $B \to \phi(\rho) K \gamma$ in perturbative QCD approach }

\author{Chao Wang}\email{wangchao88@ihep.ac.cn}
\author{Jing-Bin Liu}\email{liujb@ihep.ac.cn}
\affiliation{Institute of High Energy Physics, CAS, P.O. Box 918, Beijing 100049, China \\
 University of Chinese Academy of Sciences, Beijing 100049, China}
\author{Hsiang-nan Li}\email{hnli@phys.sinica.edu.tw}
\affiliation{%
Institute of Physics, Academia Sinica, Taipei, Taiwan 115, Republic of China}
\author{Cai-Dian L\"u}\email{lucd@ihep.ac.cn}
\affiliation{Institute of High Energy Physics, CAS, P.O. Box 918, Beijing 100049, China \\
 University of Chinese Academy of Sciences, Beijing 100049, China}
\date{\today}

\begin{abstract}
We study the three-body radiative decays $B\to \phi(\rho) K\gamma$ induced by
a flavor-changing neutral current in the perturbative QCD approach.
Pseudoscalar-vector ($PV$) distribution amplitudes (DAs) are introduced for the
final-state $\phi K$ ($\rho K$) pair to capture important infrared dynamics in
the region with a small $PV$-pair invariant mass. The dependence of these $PV$ DAs
on the parton momentum fraction is parametrized in terms of the Gegenbauer
polynomials, and the dependence on the meson momentum fraction is derived
through their normalizations to time-like $PV$ form factors. In addition to the dominant
electromagnetic penguin, the subleading chromomagnetic penguin,
quark-loop and annihilation diagrams are also calculated. After determining the
$PV$ DAs from relevant branching-ratio data, the direct $CP$ asymmetries and decay spectra
in the $PV$-pair invariant mass are predicted for each $B\to \phi(\rho) K\gamma$
mode.
\end{abstract}

\maketitle

\section{\label{sec-1}Introduction}

A large number of experimental investigations on three-body hadronic
$B$ meson decays have been carried out by the BABAR~\cite{Lees:2012kxa,
Lees:2011nf,Aubert:2009me,BABAR:2011ae,Aubert:2009av,Aubert:2008bj},
Belle~\cite{Garmash:2004wa,Garmash:2006fh,Chang:2004um,Garmash:2005rv}
and LHCb~\cite{Aaij:2013sfa,Aaij:2013bla,Aaij:2014iva} Collaborations in
recent decades. The forthcoming Belle-II experiments will further
improve the accuracy of their measurements by means of Dalitz
analyses~\cite{Aushev:2010bq}. These decays, involving QCD dynamics much
more complicated than in two-body cases, impose a severe challenge to the
development of corresponding theoretical frameworks. The currently
available frameworks based on the factorization theorem for three-body
hadronic $B$ meson decays include the perturbative QCD (PQCD)
approach~\cite{Chen:2002th,Wang:2014ira,Wang:2015uea,Wang:2016rlo} and the
QCD-improved factorization approach~\cite{Wei:2003ti,ElBennich:2009da,Krankl:2015fha}.
Though the factorization theorem is not yet proved rigorously,
phenomenological applications have been attempted, and abundant predictions
have been made. There exist other approaches, such as final state
interactions~\cite{Furman:2005xp} and heavy meson chiral perturbation
theory~\cite{Cheng:2002qu,Cheng:2007si}. The
stringent confrontation of theoretical predictions with data in the whole
final state phase space is likely to reveal new dynamics, signifying
the importance of three-body hadronic $B$ meson decays.

Most of PQCD studies of the above decays
focus on the kinematic configuration corresponding to edges of
Dalitz plots, whose formalism can be simplified by the introduction of
two-hadron distribution amplitudes (DAs)~\cite{Chen:2002th}. In these
regions two of the three final state hadrons collimate with each other in
the rest frame of the $B$ meson. At the quark level this configuration
involves the hadronization of two energetic collinear quarks, produced
from the $b$ quark decay, into the two collimated hadrons. The
hadron-pair system, dominated by infrared QCD dynamics, can then be factorized
out of the whole process, and defines the two-hadron DA
$\Phi_{h_1h_2}$~\cite{Diehl:1998dk,Polyakov:1998ze,Mueller:1998fv,Diehl:2000uv}.
The factorization formula for the $B\to h_1h_2h_3$ decay is then expressed as
\begin{eqnarray}
\mathcal{M}=\Phi_B\otimes H\otimes\Phi_{h_1h_2}\otimes\Phi_{h_3},
~\label{eq1}
\end{eqnarray}
where $\Phi_B$ ($\Phi_{h_3}$) denotes the $B$ meson ($h_3$ hadron) DA, and
$\otimes$ means the convolution in
parton momenta. The hard kernel $H$ for the $b$ quark decay, similar to the
two-body case, starts with the diagrams of single hard gluon exchange. An advantage
of the above formalism is that both resonant and nonresonant contributions to
the hadron-pair system can be included into the two-hadron DA through appropriate
parametrization. Although Eq.~(\ref{eq1}) has been applied to the whole
three-body phase space, it should be understood that it is precise only in the
region with a small hadron-pair invariant mass. A two-hadron DA loses its accuracy
in the central region of a Dalitz plot, where the major contribution to three-body
decays arises from two hard gluon exchanges~\cite{Chen:2002th}. Nevertheless,
it is also the region, where a two-hadron DA decreases with certain power law of
the invariant mass, and gives a minor contribution.

In this paper we will extend the PQCD approach to the three-body
radiative decays $B\to PV\gamma$ with $P$ ($V$) representing a pseudoscalar
(vector) meson. The significance of these decays has been well recognized: the involved
flavor-changing neutral current $b\to s\gamma$, occurring only at loop
level in the Standard Model, is sensitive to new physics effects.
Following the similar reasoning, the two-hadron DAs $\Phi_{PV}$ can be
introduced to collect the dominant contribution from the region with
a small $PV$-pair invariant mass $m_{PV}$. For instance, nearly $72\%$ of the
signal events appear in the low mass region with
$m_{\phi K}\in[1.5,2.0]$ GeV~\cite{Sahoo:2011zd}. Besides, the emitted
photons from the leading electromagnetic $O_{7\gamma}$ transition are
mainly left-handed (right-handed) in $B^-$ and $\bar{B}^0$ ($B^+$ and $B^0$)
meson decays. A chirality flip may be induced by local four-quark operators
and the chromomagnetic penguin operator $O_{8g}$ from QCD corrections,
as well as by final state interactions among various resonant
channels~\cite{Gronau:2017kyq}. We will address the above subjects,
taking the $B\to \phi(\rho) K\gamma$ decays as examples.
The resonant contribution to the $\phi K$ system is
negligible~\cite{Sahoo:2011zd,Aubert:2006he}, so the
parametrization for the DAs $\Phi_{\phi K}$ in Ref.~\cite{Chen:2004az},
which contain time-like form factors with certain power-law behavior, is adopted.
The resonant contributions from the states $K_1(1270)$ and $K^*(1680)$ to the
$\rho K$ system dominate~\cite{Nishida:2002me,Sanchez:2015pxu}. Therefore,
the parametrization of the DAs $\Phi_{\rho K}$ follows that for quasi-two-body
$B$ meson decays~\cite{Wang:2015uea,Wang:2016rlo,Li:2016tpn,Ma:2017idu},
namely, the Breit-Wigner model.

In Sec.~\ref{sec-2} we construct the
$PV$ DAs according to the procedure proposed in~\cite{Chen:2004az}:
the dependence on the parton momentum fraction is parametrized in terms of
the Gegenbauer polynomials, and the dependence on the meson momentum fraction
is derived through the normalizations to time-like $PV$ form factors.
In Sec.~\ref{sec-3} we analyze the three-body
radiative decays $B\to \phi(\rho) K\gamma$, determine the $PV$ DAs from
relevant branching-ratio data, and then predict their direct $CP$ asymmetries,
photon polarization asymmetries, and decay spectra in the $PV$-pair
invariant mass. Our work is more complete than~\cite{Chen:2004az},
because the contributions from the operators $O_{7\gamma}$, $O_{8g}$,
and $O_2$ are all considered, and the annihilation diagrams are calculated.
The summary is given in the last Section, and the factorization formulas
for the $B\to \phi(\rho) K\gamma$ decay amplitudes are collected in the Appendix.


\section{\label{sec-2}Pseudoscalar-Vector distribution amplitudes}

We choose the $B$ meson momentum $P_B$, the $PV$-pair momentum $P$,
and the photon momentum $P_\gamma$ in the light-cone coordinates as
\begin{eqnarray}
P_B=\frac{m_B}{\sqrt{2}}(1,1,\bm{0}_T), \qquad
P=\frac{m_B}{\sqrt{2}}(1,\eta,\bm{0}_T),
\quad P_\gamma=\frac{m_B}{\sqrt{2}}(0,1-\eta,\bm{0}_T),\label{mom}
\end{eqnarray}
with the $B$ meson mass $m_B$ and the variable $\eta=P^2/m_B^2\equiv \omega^2/m_B^2$,
$\omega$ being the invariant mass of the $PV$ pair. Define the momenta
of the vector and pseudoscalar mesons by
\begin{eqnarray}
P_1=&&(\zeta P^+,  [(1-\zeta)\eta+r_V^2]P^+,
\sqrt{(\zeta\omega^2-m_V^2)(1-\zeta)}, 0) , \nonumber \\
P_2=&&((1-\zeta)P^+,  (\zeta\eta-r_V^2)P^+, -\sqrt{(\zeta\omega^2-m_V^2)(1-\zeta)}, 0),
\end{eqnarray}
respectively, which obey $P=P_1+P_2$, with the vector meson momentum fraction
$\zeta$ and the mass ratio $r_V=m_V/m_B$. The smaller pseudoscalar mass has been neglected.
Write the spectator momenta in the $B$ meson and in the
$PV$ pair as
\begin{eqnarray}
k_1=(0, \frac{m_B}{\sqrt{2}}x_1, \bm{k}_{1T}), \quad k_2=(\frac{m_B}{\sqrt{2}}z, 0, \bm{k}_{2T}),
\end{eqnarray}
respectively, $x_1$ and $z$ being the momentum fractions.
We also define the polarization vectors $\epsilon$ of the $PV$ system by
\begin{eqnarray}
\epsilon^*(\pm)=\frac{1}{\sqrt{2}}(0,0,\mp1,-i),
\quad \epsilon_L^*=\frac{1}{\sqrt{2\eta}}(1,-\eta,\bm{0}_T).
\end{eqnarray}

A two-meson DA $\phi(z,\zeta,\omega)$ describes the hadronization of two
collinear quarks, together with other quarks popped out of the
vacuum and playing no role in a hard decay process, into two collimated mesons.
It can be decomposed in terms of the
eigenfunctions of the ERBL evolution equation~\cite{Lepage:1979zb,Efremov:1979qk},
i.e., the Gegenbauer polynomials $C_{n}^{3/2}(2z-1)$, and the partial waves, i.e.,
the Legendre polynomials $P_l(2\zeta-1)$~\cite{Polyakov:1998ze,Diehl:2000uv}.
However, for the $PV$ system, the different spins of the pseudosalar and the vector
render the Legendre polynomial expansion not applicable. Hence, we
extract the $\zeta$ dependence from the normalizations of the $PV$ DAs to the
associated time-like form factors~\cite{Chen:2004az}, which depend on the
$PV$-pair invariant mass $\omega$, a procedure
similar to deriving the two-pion DAs via the process
$\gamma\gamma^*\to\pi^+\pi^-$~\cite{Diehl:1999ek}.
To be explicit, we evaluate perturbatively the matrix elements
of local currents
\begin{eqnarray}
\langle V(P_1,\epsilon^*(V))P(P_2)|\bar{q}'(0)\Gamma q(0)|0\rangle,
\end{eqnarray}
using the vector and pseudoscalar DAs up to twist 3,
where the polarization vectors of the vector meson satisfy
$\epsilon^*(V)\cdot P_1=0$ and $\epsilon^*(V)^{2}=-1$, and
$\Gamma$ represents the possible spin projectors $I$, $\gamma_5$,
$\gamma_\mu$, $\gamma_\mu\gamma_5$, and $\sigma_{\mu\nu}\gamma_5$. The above
matrix element is precisely the normalization of the $PV$ DA associated with the
spin projector $\Gamma$, and also the $PV$ time-like form factor associated
with the local current $\bar{q}'\Gamma q$. The goal of the perturbative
calculation is to reveal the kinematic structure of the matrix element in
terms of $P_1$, $\epsilon(V)$, and $P_2$ for
each $\Gamma$, which are then approximated by the momentum $P$
and the polarization vectors $\epsilon$ of the $PV$ system according to the
power counting rules in the heavy quark limit~\cite{Chen:2004az}.
In this way we obtain the $\zeta$ dependence of the $PV$ DAs up to twist 3,
i.e., $\mathcal{O}(\omega/m_B)$.

The expansions of the nonlocal matrix elements for various spin projectors $\Gamma$
up to twist 3 are listed below:
\begin{eqnarray}
\langle PV|\bar{q}'(y^-)\gamma_\mu\gamma_5 q(0)|0\rangle
&=&P_\mu \int_0^1dze^{izP\cdot y}\phi_\parallel(z,\zeta,\omega)
+\omega\epsilon^*_{T\mu}\int_0^1dze^{izP\cdot y}\phi_a(z,\zeta,\omega)\label{eq-m5}, \\
\langle PV|\bar{q}'(y^-)\sigma_{\mu\nu}\gamma_5q(0)|0\rangle
&=&-i\big\{(\epsilon^*_{T\mu}P_\nu-\epsilon^*_{T\nu}P_{\mu})
\int_0^1dze^{izP\cdot y}\phi_T(z,\zeta,\omega) \nonumber \\
&&+  (\epsilon^*_{L\mu}P_\nu-\epsilon^*_{L\nu}P_{\mu})
\int_0^1dze^{izP\cdot y}\phi_3(z,\zeta,\omega)\big\}\label{eq-smn},\\
\langle PV|\bar{q}'(y^-)\gamma_5 q(0)|0\rangle &=&
\omega\int_0^1e^{izP\cdot y}\phi_p(z,\zeta,\omega),\\
\langle PV|\bar{q}'(y^-)\gamma_\mu q(0)|0\rangle
&=&i\frac{\omega}{P\cdot n_-}\epsilon_{\mu\nu\rho\sigma}
\epsilon_T^{*\nu} P^\rho n_-^\sigma \int_0^1e^{izP\cdot y}\phi_v(z,\zeta,\omega)\label{eq-m},\\
\langle PV|\bar{q}'(y^-)Is(0)|0\rangle =&0 \label{eq-1},
\end{eqnarray}
where $n_-=(0,1,\bm{0}_T)$ is a light-like vector, and the convention
$\epsilon_{0123}=-1$ has been employed. To get the first term in
Eq.~(\ref{eq-m5}), we have applied
\begin{eqnarray}
(P_1-P_2)_\mu\simeq (2\zeta-1)P_\mu,
\end{eqnarray}
where the coefficient $2\zeta-1$ is absorbed into the DA $\phi_\parallel$,
giving rise to its $\zeta$ dependence. We have also made the approximation
\begin{eqnarray}
\epsilon^*_{T\mu}(V)P_{1\nu}-\epsilon^*_{T\nu}(V)P_{1\mu}&
\simeq & \zeta(\epsilon^*_{T\mu}P_{\nu}-\epsilon^*_{T\nu}P_{\mu}),\\
\frac{2}{\omega}(P_{1\mu}P_{2\nu}-P_{1\nu}P_{2\mu})&
\simeq & (2\zeta-1)(\epsilon^*_{L\mu}P_\nu-\epsilon^*_{L\nu}P_{\mu}),
\end{eqnarray}
as arriving at Eq.~(\ref{eq-smn}). It is found, compared to~\cite{Chen:2004az},
that the term $(P_{1\mu}P_{2\nu}-P_{1\nu}P_{2\mu})$ does not generate the twist-2
contribution $(\epsilon_{T\mu}P_{\nu}-\epsilon_{T\nu}P_{\mu})$, since a
transverse momentum and a transverse polarization have different physical meanings.
Equation~(\ref{eq-m}) comes from the approximation of the kinematic factor
\begin{eqnarray}
\frac{2}{\omega}\epsilon_{\mu\nu\rho\sigma}\epsilon_T^{*\nu}(V)
p_1^\rho p_2^\sigma\simeq\frac{\omega}{P\cdot n_-} (2\zeta-1)
\epsilon_{\mu\nu\rho\sigma}\epsilon_T^{*\nu} P^\rho n_-^\sigma.
\end{eqnarray}

We summarize the $PV$ DAs for the longitudinal and transverse polarizations
from Eq.~(\ref{eq-m5})-Eq.~(\ref{eq-1}) as
\begin{eqnarray}
\langle PV(P,\epsilon^*_L)|\bar{q}'(y^-)_jq(0)_l|0\rangle
&=&\frac{1}{\sqrt{2N_c}}\int_0^1dze^{izP\cdot y}
\Big\{(\gamma_5\slashed{P})_{lj} \phi_\parallel(z,\zeta,\omega)
+(\gamma_5)_{lj}\omega\phi_p(z,\zeta,\omega)\nonumber\\
&&+(\gamma_5\slashed{\epsilon}^*_L\slashed{P})_{lj}\phi_3(z,\zeta,\omega)\Big\}, \nonumber \\
\langle PV(P,\epsilon^*_T)|\bar{q}'(y^-)_jq(0)_l|0\rangle
&=&\frac{1}{\sqrt{2N_c}}\int_0^1dze^{izP\cdot y}
\Big\{(\gamma_5\slashed{\epsilon}^*_T\slashed{P})_{lj}
\phi_t(z,\zeta,\omega)+(\gamma_5\slashed{\epsilon}^*_{T\mu})_{lj}\omega
\phi_a(z,\zeta,\omega) \nonumber \\
&&+i\frac{\omega}{P\cdot n_-}\epsilon_{\mu\nu\rho\sigma}
(\gamma^\mu)_{lj}\epsilon_T^{*\nu} P^\rho n_-^\sigma \phi_v(z,\zeta,\omega)\Big\},
\end{eqnarray}
where $\phi_{\parallel,t}$ are of twist 2, and $\phi_{p,3,a,v}$ are of twist 3.
The above $PV$ DAs contain the products of the time-like
form factors $F(\omega)$, which define the normalizations of the DAs, and
the $z$-dependent and $\zeta$-dependent functions:
\begin{eqnarray}
\phi_\parallel(z,\zeta,\omega)&=&\frac{3F_\parallel(\omega)}{\sqrt{2N_c}}f_\parallel(z)(2\zeta-1), \nonumber\\
\phi_p(z,\zeta,\omega)&=&\frac{3F_p(\omega)}{\sqrt{2N_c}}f_p(z), \nonumber\\
\phi_3(z,\zeta,\omega)&=&\frac{3F_3(\omega)}{\sqrt{2N_c}}f_3(z)(2\zeta-1), \nonumber\\
\phi_t(z,\zeta,\omega)&=&\frac{3F_T(\omega)}{\sqrt{2N_c}}f_t(z)\zeta, \nonumber\\
\phi_a(z,\zeta,\omega)&=&\frac{3F_a(\omega)}{\sqrt{2N_c}}f_a(z),\nonumber\\
\phi_v(z,\zeta,\omega)&=&\frac{3F_v(\omega)}{\sqrt{2N_c}}f_v(z)(2\zeta-1)\label{fzz}.
\end{eqnarray}
Different from Ref.~\cite{Chen:2004az}, the DA $\phi_a$ in the above expressions
does not depend on the meson momentum fraction $\zeta$.
Note that only the DAs for the transversely polarized $PV$ pair are relevant
to the three-body radiative decays $B\to PV\gamma$ considered here.

We include the first Gegenbauer moment for the function $f_a(z)$, making
the $\phi K$ DA $\phi_a$ a bit asymmetric in the parton momentum distribution,
and assume the asymptotic form $z(1-z)$ for the functions $f_{t,v}(z)$
for simplicity,
\begin{eqnarray}
\phi_t(z,\zeta,\omega)&=&\frac{3F_T^{\phi K}(\omega)}{\sqrt{2N_c}}z(1-z)\zeta, \nonumber\\
\phi_a(z,\zeta,\omega)&=&\frac{3F_a^{\phi K}(\omega)}{\sqrt{2N_c}}z(1-z)
\left[1+a_1C_1^{3/2}(2z-1)\right],\nonumber\\
\phi_v(z,\zeta,\omega)&=&\frac{3F_v^{\phi K}(\omega)}{\sqrt{2N_c}}z(1-z)(2\zeta-1).
\label{gen}
\end{eqnarray}
The $\phi K$ time-like form factors, dominated by nonresonant contributions,
are parametrized as~\cite{Chen:2004az}
\begin{eqnarray}
F_T^{\phi K}(\omega)&=&\frac{m_T^2}{(\omega-m_l)^2+m_T^2}, \nonumber \\
F_a^{\phi K}(\omega)&=&F_v^{\phi K}(\omega)
=\frac{m_0m_\parallel^2}{(\omega-m_l)^3+m_0m_\parallel^2} \label{ff1},
\end{eqnarray}
with the chiral scale $m_0\simeq1.7$ GeV~\cite{Huang:2004tp} and the
threshold invariant mass $m_l=m_\phi+m_K$. That is, we keep the pseudoscalar
mass only in the phase space allowed for the time-like form factors.
The tunable parameters $a_1$ and $m_T\simeq m_\parallel$, expected to be few GeV,
will be determined from the fit to the data of the $B\to\phi K\gamma$ branching ratios.
Since $\phi_a$ gives a larger contribution, as verified numerically
in the next section, the data lead stronger constraint to its
first Gegenbauer moment. This explains why we introduce $a_1$ only into $\phi_a$.

The amplitude analysis on the resonant structure of the final state
in the $B^+\to  K^+\pi^-\pi^+\gamma$ decay~\cite{Sanchez:2015pxu} provides a
useful guideline for parametrizing the resonant contribution to the $B\to\rho K\gamma$
mode, for which $K_1(1270)$ and $K^*(1680)$ are the major intermediate resonances.
The resonance $K_1(1270)$ is a mixture of the $K_{1A}(1^3P_1)$ and
$K_{1B}(1^1P_1)$ states,
\begin{eqnarray}
K_1(1270) &=& \sin\theta_{K} K_{1A} + \cos\theta_{K} K_{1B}, \nonumber \\
K_1(1400) &=& \cos\theta_{K} K_{1A} - \sin\theta_{K} K_{1B},\label{kmix}
\end{eqnarray}
$\theta_K$ being the mixing angle.
With the insertion of Eq.~(\ref{kmix}), the quasi-two-body
$B\to K_1(1270)(\to\rho K) \gamma$ decay amplitude can be expressed as the
combination of the $B\to K_{1A}(\to\rho K) \gamma$ amplitude and
the $B\to K_{1B}(\to\rho K) \gamma$ amplitude, such that
the $K_{1A}$ and $K_{1B}$ meson DAs with the specific symmetry in the
$z$ dependence can be employed. The $\rho K$ DAs
for the $B\to K_1(1270)(\to \rho K)\gamma$ modes are then
parametrized as
\begin{eqnarray}
\phi_t(z,\zeta,\omega)&=&\frac{3F_1^{\rho K}(\omega)}{\sqrt{2N_c}}c_K
z(1-z)\left[a_0^\perp+3a_1^\perp (2z-1)+a_2^\perp\frac{3}{2}(5(2z-1)^2-1)\right]\zeta, \nonumber\\
\phi_a(z,\zeta,\omega)&=&\frac{F_1^{\rho K}(\omega)}{2\sqrt{2N_c}}c_K
\left[\frac{3}{4}a_0^\parallel(1+(2z-1)^2)+\frac{3}{2}a_1^\parallel (2z-1)^3\right],\nonumber\\
\phi_v(z,\zeta,\omega)&=&\frac{3F_1^{\rho K}(\omega)}{4\sqrt{2N_c}}c_K
\left[a_0^\parallel (1-2z)+a_1^\parallel (6z-6z^2-1)\right](2\zeta-1),
\end{eqnarray}
where the mixing factor $c_K$ and the Gegenbauer moments
associated with the $K_{1A}(K_{1B})$ state take the values
\begin{eqnarray}
c_K&=&\sin\theta_K\;(\cos\theta_K),
a_0^\parallel=1(-0.15\pm0.15), a_1^\parallel=-0.30_{-0.00}^{+0.26}(-1.95+0.45),\nonumber\\
a_0^\perp&=&0.26_{-0.22}^{+0.03}(1), a_1^\perp=-1.08\pm0.48(0.30_{-0.31}^{+0.00}),
a_2^\perp=0.02\pm0.21(-0.02\pm0.22).
\end{eqnarray}
In principle, the Gegenbauer moments for the $\rho K$ DAs are free
parameters. Here we adopt those for the $K_{1A}$ and $K_{1B}$
DAs~\cite{Yang:2007zt,Cheng:2008gxa} as their typical values in
the numerical study below. The form factor $F_1^{\rho K}$ picks up the
standard Breit-Winger model,
\begin{eqnarray}
F_1^{\rho K}(\omega)&=&\frac{m_{K_1(1270)}^2 }{m^2_{K_1(1270)}-\omega^2-im_{K_1(1270)}\Gamma_{K_1(1270)}},
\end{eqnarray}
$m_{K_1(1270)}$ ($\Gamma_{K_1(1270)}$) being the mass (decay width) 
of the $K_1(1270)$ meson.

In this paper we have proposed different parametrizations of the nonresonant and
resonant contributions: for the former, the final state interaction is
ignored, and their form factors are real and normalized to $F(m_{l})=1$.
Because it arises from a wider range of the invariant mass, the different
power-law behaviors of the form factors, $F_{T}^{\phi K}\sim 1/\omega^2$ and
$F_{a,v}^{\phi K}(\omega) \sim m_0/\omega^3$
at large $\omega$~\cite{Keum:2000ph,Keum:2000wi,Keum:2000ms},
have been taken into account. For the latter, the major piece comes from the region
around the resonance mass, so it is reasonable not to differentiate the power-law
behaviors of the form factors in $\phi_{t,a,v}$, but parametrize them in terms
of the same Breit-Wigner model.

Another resonance $K^*(1680)$ contributes via the
$B\to K^*(1680)(\to \rho K)\gamma $ channel~\cite{Nishida:2002me,Sanchez:2015pxu}.
Since the Gegenbauer moments of the $K^*(1680)$ DAs are unknown, we simply assume
the asymptotic form $z(1-z)$ for the $z$ dependence of the corresponding
$\rho K$ DAs. The standard Breit-Wigner model is aslo used for
the associated form factor
\begin{eqnarray}
F_{2}^{\rho K}(\omega)=\frac{cm_{K^*(1680)}^2}{m^2_{K^*(1680)}
-\omega^2-im_{K^*(1680)}\Gamma_{K^*(1680)}},\label{ff2}
\end{eqnarray}
where the parameter $c$, characterizing the strength relative to the amplitude
from the resonance $K_1(1270)$, will be determined from the fit to
the data of the $B\to\rho K\gamma$ branching ratios. There is no interference
between the $K_1$ and $K^*$ states due to different quantum numbers. Denoting the
amplitude from the $K_1(1270)$ ($K^*(1680)$) channel by $\mathcal{A}_1$
($\mathcal{A}_2$), we compute the total amplitude squared for the
$B\to \rho K\gamma$ decays by~\cite{Sanchez:2015pxu}
\begin{eqnarray}
|\mathcal{A}|^2=\frac{1}{1+c^2}[|\mathcal{A}_1|^2+|\mathcal{A}_2|^2],\label{ff3}
\end{eqnarray}
in which the factor $1/(1+c^2)$ plays the role of an overall normalization.

\section{\label{sec-3}Numerical results}

As stated before, the evaluation of the three-body radiative
$B$ meson decay amplitudes reduces to that of two-body
ones~\cite{Keum:2004is,Lu:2005yz,Wang:2007an} with the introduction
of the $PV$ DAs. We consider the $O_{7\gamma}$, $O_{8g}$,
and $O_{2}$ operators, and the annihilation contributions,
performing an analysis more complete than in~\cite{Chen:2004az}, where only the
emission diagrams from $O_{7\gamma}$ were taken into account. The
explicit factorization formulas for various contributions are
collected in the Appendix.
The $B\to PV\gamma$ differential decay rate, i.e. the decay spectrum
in the $PV$ invariant mass, is then derived from
\begin{eqnarray}
\frac{d\Gamma}{d\omega}=\frac{1}{128\pi^3}\sqrt{\eta}(1-\eta)
\int_{m_{l}^2/\omega^2}^{1} d\zeta (|\mathcal{A}^R|^2+|\mathcal{A}^L|^2),
\end{eqnarray}
where the vector meson momentum fraction $\zeta$ is bounded by 
$m_{l}^2/\omega^2\leq\zeta\leq1$, and $\mathcal{A}^{R(L)}$
denotes the amplitude for the right-handed (left-handed)
photon emission.

The inputs for the masses (in units of GeV)~\cite{Olive:2016xmw},
the widths of the $K_1$ and $K^*$ mesons (in units of GeV)~\cite{Sanchez:2015pxu},
and the mean lifetimes of the $B$ mesons (in units of ps) are listed below:
\begin{eqnarray}
&&m_{B^{\pm,0}}=5.280, \quad m_{\phi}=1.019, \quad m_{K^{\pm}}=0.494,
\quad m_{K^0}=0.498, \quad m_{\rho}=0.775, \nonumber\\
&&m_{K_1(1270)} = 1.272, \quad m_{K^*(1680)}=1.717,
\quad \Gamma_{K_1(1270)} = 0.098 \quad \Gamma_{K^*(1680)}=0.377,\nonumber\\
&&\tau_{B^\pm}=1.638, \quad \tau_{B^0}=1.519.
\end{eqnarray}
Phenomenological investigations in the literature and experimental data
have indicated that the mixing angle $\theta_K$ for the $K_{1A}$ and
$K_{1B}$ mesons is around either $33^\circ$ or
$58^\circ$~\cite{Suzuki:1993yc,Burakovsky:1997dd,Cheng:2003bn,
Yang:2010ah,Hatanaka:2008xj,Tayduganov:2011ui,Divotgey:2013jba}.
As to the Cabibbo-Kobayashi-Maskawa (CKM) matrix elements,
we employ the Wolfenstein parametrization
with the inputs~\cite{Olive:2016xmw},
\begin{eqnarray}
\lambda&=&0.222506\pm0.00050, \quad A=0.811\pm0.026,\quad
\bar{\rho}=0.124_{-0.018}^{+0.019}, \quad \bar{\eta}=0.356\pm0.011. \nonumber
\end{eqnarray}
In addition, we take the $B$ meson decay constant $f_{B}=0.190$ GeV, which is in
agreement with the lattice results $f_{B}=0.186\pm0.004$ GeV~\cite{Dowdall}
and $f_{B}=0.186\pm0.013$ GeV~\cite{Bernardoni}, and with those from
the recent theoretical studies~\cite{Yang,Baker}.

We consider the following theoretical errors. The first source
of errors originates from the hadronic parameters,
specifically the shape parameter of the $B$ meson DA,
$\omega_B=0.40\pm0.04$ GeV. For the $B\to \rho K\gamma$ decays, this
source of errors also includes the variation of the Gegenbauer moments
in the $\rho K$ DAs. The second source characterizes the next-to-leading-order
effects in the PQCD approach: we vary the hard scale $t$ from $0.80t$ to $1.2t$
(without changing $1/b_i$) and the QCD scale
$\Lambda_{\rm QCD}=0.25\pm0.05$ GeV. The CKM matrix elements $V_{tb}$ and $V_{ts}$
involved in the dominant operator $O_{7\gamma}$ have small uncertainties, whose
errors are ignored in our numerical analysis.

The Gegenbauer moment $a_1$ in Eq.~(\ref{gen}) and the free parameters
$m_{T,\parallel}$ in Eq.~(\ref{ff1}) can be extracted from the branching-ratio
data~\cite{Olive:2016xmw}
\begin{eqnarray}
\mathcal{B}_{\rm exp}(B^+ \to \phi K^+ \gamma)&=&(2.7\pm0.4)\times 10^{-6},\nonumber \\
\mathcal{B}_{\rm exp}(B^0 \to  \phi K^0 \gamma)&=&(2.7\pm0.7)\times 10^{-6}.
\end{eqnarray}
We obtain for $a_1=-0.3$ and $m_{T,\parallel}= 3.0$ GeV,
\begin{eqnarray}
\mathcal{B}(B^+ \to \phi K^+ \gamma)&=&(2.69_{-0.18-0.36}^{+0.18+0.43})\times 10^{-6},\nonumber \\
\mathcal{B}(B^0 \to  \phi K^0 \gamma)&=&(2.41_{-0.18-0.35}^{+0.14+0.37})\times 10^{-6},\label{phik}
\end{eqnarray}
which match the observed values well. The negative $a_1$ implies that
the light spectator quark intends to carry, as expected, a smaller fraction of 
the $\phi K$ pair momentum. It has been examined that
the above results are stable against the variation of $m_{T,\parallel}$ around
few GeV. The central value of $\mathcal{B}(B^+ \to \phi K^+ \gamma)$ in Eq.~(\ref{phik})
is lower than the prediction $(2.9^{+0.7}_{-0.5})\times 10^{-6}$ in~\cite{Chen:2004az}
because of the combined effects of the following changes: retaining the kaon mass here
suppresses the phase space, and the inclusion of the parton $k_T$ renders hard propagators
more off-shell, but the asymmetric DA $\phi_a$ compensates the above reduction a bit.

The exclusive $B$ meson decays into the resonances $K_1(1270)$ and $K^*(1680)$ have
been reported by BaBar with the branching ratios~\cite{Sanchez:2015pxu},
\begin{eqnarray}
\mathcal{B}_{exp}(B^+\to K_1(1270)^+\gamma)&=&(44.1_{-7.3}^{+8.6})\times 10^{-6}, \nonumber \\
\mathcal{B}_{exp}(B^+\to K^*(1680)^+\gamma)&=&(66.7_{-13.8}^{+17.1})\times 10^{-6}.
\end{eqnarray}
The branching factions of the $K_1(1270)$ and $K^*(1680)$ transitions into the
$\rho K$ final state~\cite{Olive:2016xmw}
\begin{eqnarray}
\mathcal{B}_{\rm exp}(K_1(1270)\to \rho K)&=&(42\pm6)\%, \nonumber \\
\mathcal{B}_{\rm exp}(K^*(1680)\to \rho K)&=&(31.4_{-2.1}^{+5.0})\%,
\end{eqnarray}
then lead to
\begin{eqnarray}
\mathcal{B}_{\rm exp}(B^+\to K_1(1270)^+(\to \rho^0 K^+)\gamma)
&=&(6.2_{-1.3}^{+1.5}) \times10^{-6}, \label{sum1}\\
\mathcal{B}_{\rm exp}(B^+\to K^*(1680)^+(\to \rho^0K^+)\gamma)
&=&(7.0_{-1.5}^{+2.1}) \times10^{-6}.\label{sum2}
\end{eqnarray}
Note
that the total branching ratio~\cite{Sanchez:2015pxu}
\begin{eqnarray}
\mathcal{B}_{\rm exp}(B^+\to\rho^0 K^+\gamma) =(8.2\pm0.9)\times10^{-6},\label{sum3}
\end{eqnarray}
deviates from the sum of Eqs.~(\ref{sum1}) and (\ref{sum2}), since
the nonresonant contribution and the other minor $K_1(1400)$,
$K^*(1410)$, and $K^*_2(1430)$ resonant contributions, which may cause destructive
interference, have been neglected.
The free parameter $c$ in Eq.~(\ref{ff2}), characterizing the magnitude of the
$K^*(1680)$ resonant contribution relative to the $K_1(1270)$ one, will be determined
by the fit to Eqs.~(\ref{sum1}) and (\ref{sum2}). Note that the parameter $c$ has
reflected the strength of the $K^*(1680) \to \rho K$ transition.

\begin{figure*}
\includegraphics[width=8cm]{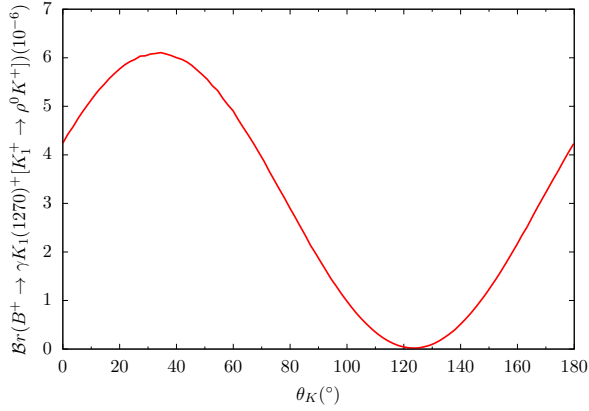}
\caption{\label{theatk}$\theta_K$ dependence of the $B^+\to K_1(1270)(\to \rho^0K^+)\gamma$
branching ratio.}
\end{figure*}

For $\theta_K=33^\circ$, the best fit value $c=2.0$ yields the results
\begin{eqnarray}
\mathcal{B}(B^+\to K_1(1270^+)(\to \rho^0 K^+)\gamma)
&=&6.11_{-1.44-1.40}^{+1.94+0.83} \times10^{-6}, \nonumber \\
\mathcal{B}(B^+\to K^*(1680^+)(\to \rho^0K^+)\gamma)
&=&6.72_{-1.63-1.41}^{+2.12+1.03}  \times10^{-6},\label{pk1}
\end{eqnarray}
and the branching ratios of the $B^0$ meson decays are predicted to be
\begin{eqnarray}
\mathcal{B}(B^0\to K_1(1270^0)(\to \rho^0 K^0)\gamma)
&=&5.00_{-1.27-1.21}^{+1.71+0.73} \times10^{-6}, \nonumber \\
\mathcal{B}(B^0\to K^*(1680^0)(\to \rho^0K^0)\gamma)
&=&6.13_{-1.53-1.32}^{+1.96+0.99} \times10^{-6}.
\end{eqnarray}
For $\theta_K=58^\circ$, we choose the best fit value $c=1.8$, obtaining
\begin{eqnarray}
\mathcal{B}(B^+\to K_1(1270^+)(\to \rho^0 K^+)\gamma)
&=&5.95_{-1.43-1.26}^{+1.92+0.75}  \times10^{-6}, \nonumber \\
\mathcal{B}(B^+\to K^*(1680^+)(\to \rho^0K^+)\gamma)
&=&6.48_{-1.55-1.34}^{+1.98+1.03} \times10^{-6},\label{pk2}
\end{eqnarray}
and predicting
\begin{eqnarray}
\mathcal{B}(B^0\to K_1(1270^0)(\to \rho^0 K^0)\gamma)
&=&4.92_{-1.24-1.09}^{+1.68+0.64} \times10^{-6}, \nonumber \\
\mathcal{B}(B^0\to K^*(1680^0)(\to \rho^0K^0)\gamma)
&=&5.84_{-1.45-1.27}^{+1.86+0.96} \times10^{-6}.
\end{eqnarray}
We mention that an upper bound $\omega\leq 1.8$ GeV for the $\rho K$ invariant
mass, the same as the experimental cutoff~\cite{Sanchez:2015pxu},
has been applied to the calculation. To test
the sensitivity of the branching ratios to the mixing angle $\theta_K$, we
fix $c=2.0$, and display the $\theta_K$ dependence of the
$B^+\to K_1(1270^+)(\to \rho^0K^+)\gamma $ branching ratio in
Fig.~\ref{theatk}.
The coordinates $\theta_K\sim33^\circ$ and $58^\circ$ happen
to locate on the two sides of a peak, explaining why the results
in Eqs.~(\ref{pk1}) and (\ref{pk2}) are close to each other.

Summing the contributions from the two quasi-two-body modes
according to Eq.~(\ref{ff3}), we get
\begin{eqnarray}
\mathcal{B}(B^+\to\rho^0 K^+\gamma)&=&
\left\{
\begin{array}{l}
12.8_{-2.9}^{+1.8}\times10^{-6}, ~\text{for}~\theta_K=33^\circ ,\\
12.4_{-2.8}^{+3.0}\times10^{-6}, ~\text{for}~\theta_K=58^\circ ,
\end{array}
\right. \label{e84}\\
\mathcal{B}(B^0\to\rho^0 K^0\gamma)&=&
\left\{
\begin{array}{l}
11.1_{-2.7}^{+2.9}\times10^{-6}, ~\text{for}~\theta_K=33^\circ ,\\
10.8_{-2.5}^{+2.8}\times10^{-6}, ~\text{for}~\theta_K=58^\circ ,
\end{array}
\right. \label{e85}
\end{eqnarray}
whose theoretical uncertainties contain only those associated with
the considered resonances. The isospin symmetry then yields the estimate
\begin{eqnarray}
\mathcal{B}(B^+\to \rho^{+}K^0\gamma)&=&2\mathcal{B}(B^+\to \rho^{0}K^+\gamma), \nonumber \\
\mathcal{B}(B^0\to \rho^{-}K^+\gamma)&=&2\mathcal{B}(B^0\to \rho^{0}K^0\gamma).
\end{eqnarray}

The direct $CP$ asymmetry in the $B\to PV\gamma$ decay is define by
\begin{eqnarray}\label{def}
A_{CP}=\frac{\mathcal{B}(\bar{B}\to\bar{P}\bar{V}\gamma)-
\mathcal{B}(B\to PV\gamma)}{\mathcal{B}(\bar{B}\to\bar{P}\bar{V}\gamma)+\mathcal{B}(B\to PV\gamma)}.
\end{eqnarray}
Since the difference of the weak phases between $V_{tb}^*V_{ts}$ and
$V_{cb}^*V_{cs}$ is negligible, the dominant $O_{7\gamma}$ contribution
can induce an appreciable $CP$ asymmetry only through its interference with
the amplitudes proportional to $V_{ub}^*V_{us}$. We predict the direct
$CP$ asymmetries (in units of percentage)
\begin{eqnarray}
A_{CP}(B^+ \to \phi K^+ \gamma) &=&-3.78_{-0.1-0.3}^{+0.2+0.7}, \label{cp1}\\
A_{CP}(B^0 \to \phi K^0 \gamma) &=&-0.13_{-0.01-0.03}^{+0.01+0.02}, \\
A_{CP}(B^+\to\rho^0 K^+\gamma)&=&
\left\{
\begin{array}{l}
-2.6_{-0.1-0.2}^{+0.1+0.2}, ~\text{for}~\theta_K=33^\circ ,\\
-2.7_{-0.1-0.2}^{+0.1+0.2}, ~\text{for}~\theta_K=58^\circ ,
\end{array}
\right. \label{e90} \\
A_{CP}(B^0 \to \rho^0 K^0 \gamma) &=&
\left\{
\begin{array}{l}
-0.16_{-0.00-0.05}^{+0.00+0.04}, ~\text{for}~\theta_K=33^\circ ,\\
-0.14_{-0.00-0.05}^{+0.00+0.04}, ~\text{for}~\theta_K=58^\circ ,
\end{array}
\right. \label{e91}
\end{eqnarray}
whose errors are smaller than those of the branching fractions,
due to the cancellation of partial theoretical uncertainties
in the ratio in Eq.~(\ref{def}).
Both the Belle and BaBar Collaborations have
measured the direct $CP$ asymmetries (in units of perventage)~\cite{Sahoo:2011zd,Aubert:2006he}
\begin{eqnarray}
A_{CP}(B^+ \to \phi K^+ \gamma)=
\left\{
\begin{array}{l}
-3\pm11\pm8, \text{(Belle)},\\
-26\pm14\pm5, \text{(BaBar)},
\end{array}
\right.
\end{eqnarray}
which are consistent with our prediction in Eq.~(\ref{cp1}).

The photon polarization parameter is defined by~\cite{Atwood:1997zr},
\begin{eqnarray}
\lambda_\gamma=\frac{|\mathcal{A}(B\to PV\gamma_R)|^2-|
\mathcal{A}(B\to PV\gamma_L)|^2}{|\mathcal{A}(B\to PV\gamma_R)|^2
+|\mathcal{A}(B\to PV\gamma_L)|^2},
\end{eqnarray}
whose measurements provide a crucial test for
the Standard Model~\cite{Gronau:2002rz,Kou:2010kn}.
We find $\lambda_\gamma\simeq1$ in our framework, implying that the
left-handed contribution is tiny in both the $B\to\phi K\gamma$ and
$B\to\rho K\gamma$ modes. This result is equivalent to the
dominance of the $O_{7\gamma}$ operator in both
the nonresonant and resonant channels.
The smallness of the $O_{8g}$, $O_{2}$ and annihilation contributions
agrees with the observation made in Ref.~\cite{Matsumori:2005ax}.


\begin{figure*}
\subfigure[$B^+\to\phi K^+\gamma$]{
\includegraphics[width=7.5cm]{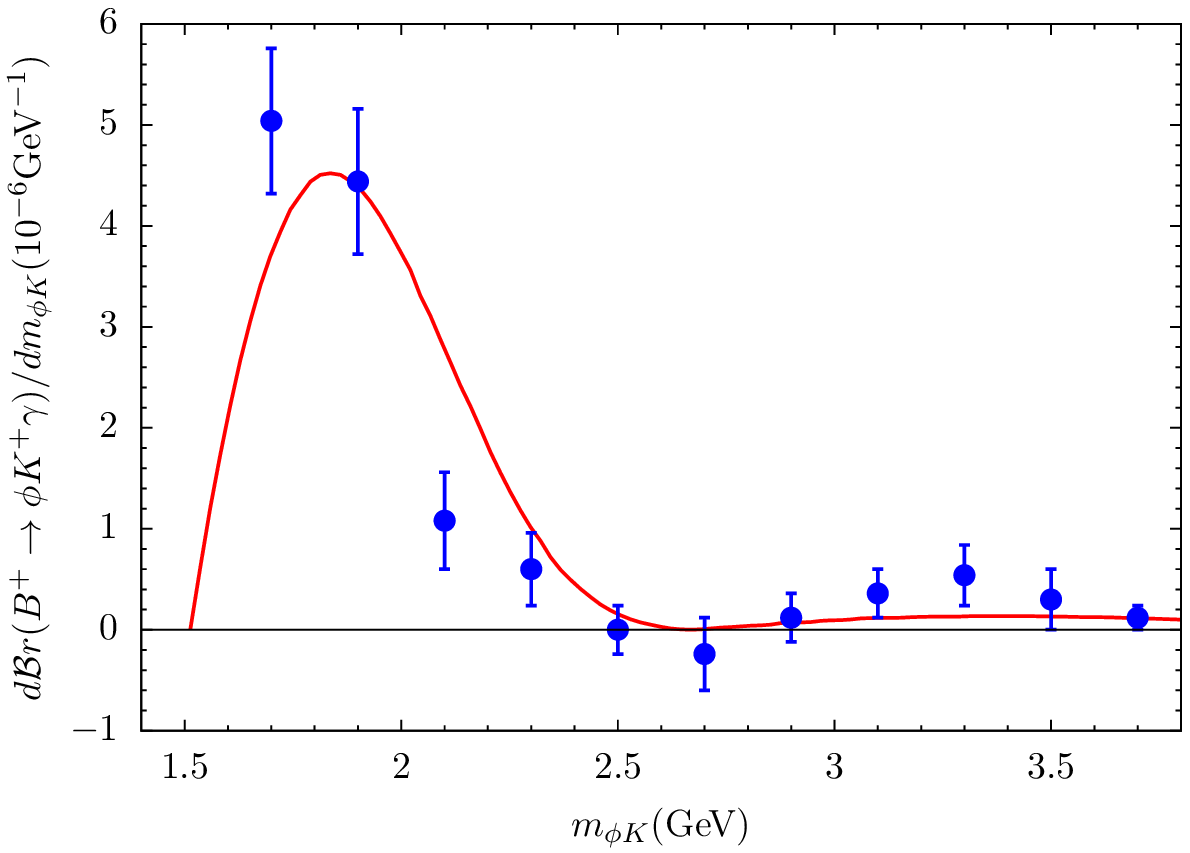}}
\subfigure[$B^0\to\phi K^0\gamma$]{
\includegraphics[width=7.5cm]{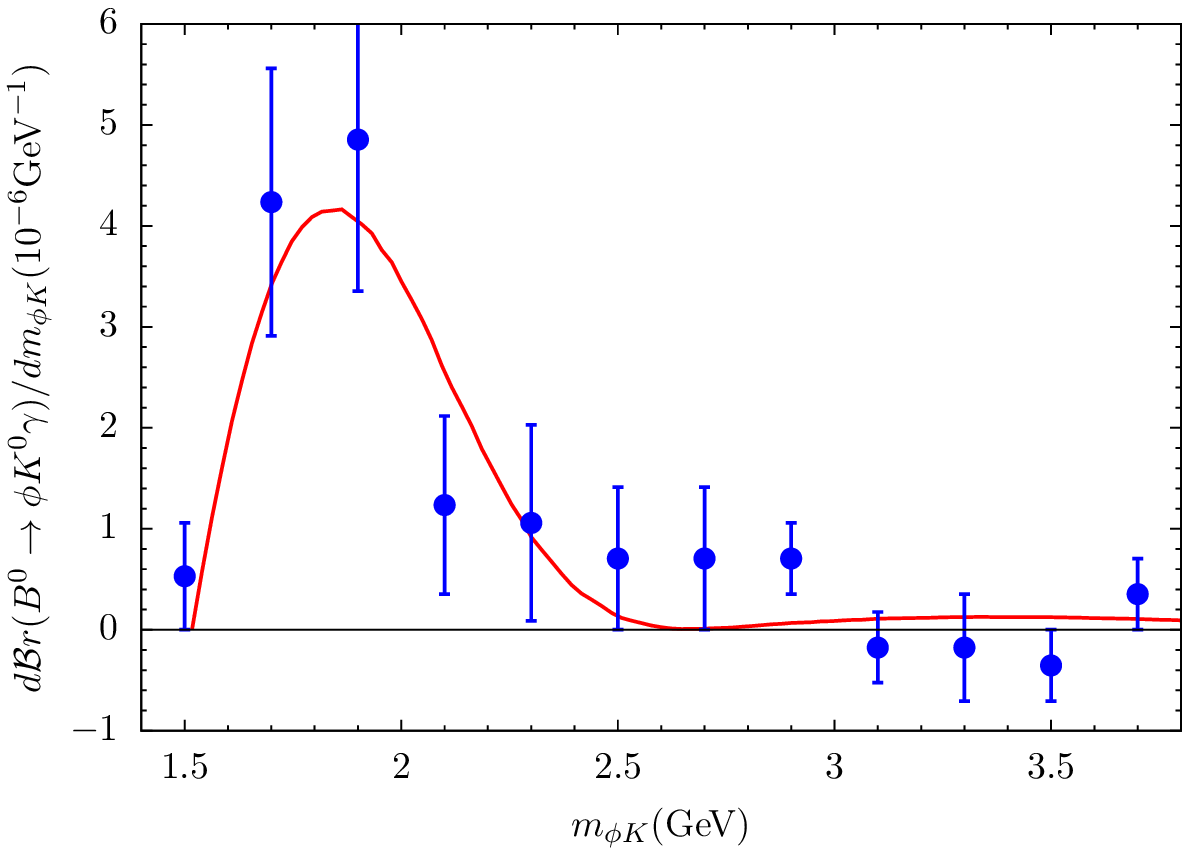}}
\caption{\label{spe1}Predicted $B\to\phi K\gamma$
decay spectra (curves) and the Belle data (points with error bars)
in the $\phi K$ invariant mass.}
\end{figure*}

\begin{figure*}
\subfigure[$B^+\to\rho^+ K^0\gamma$]{
\includegraphics[width=7.5cm]{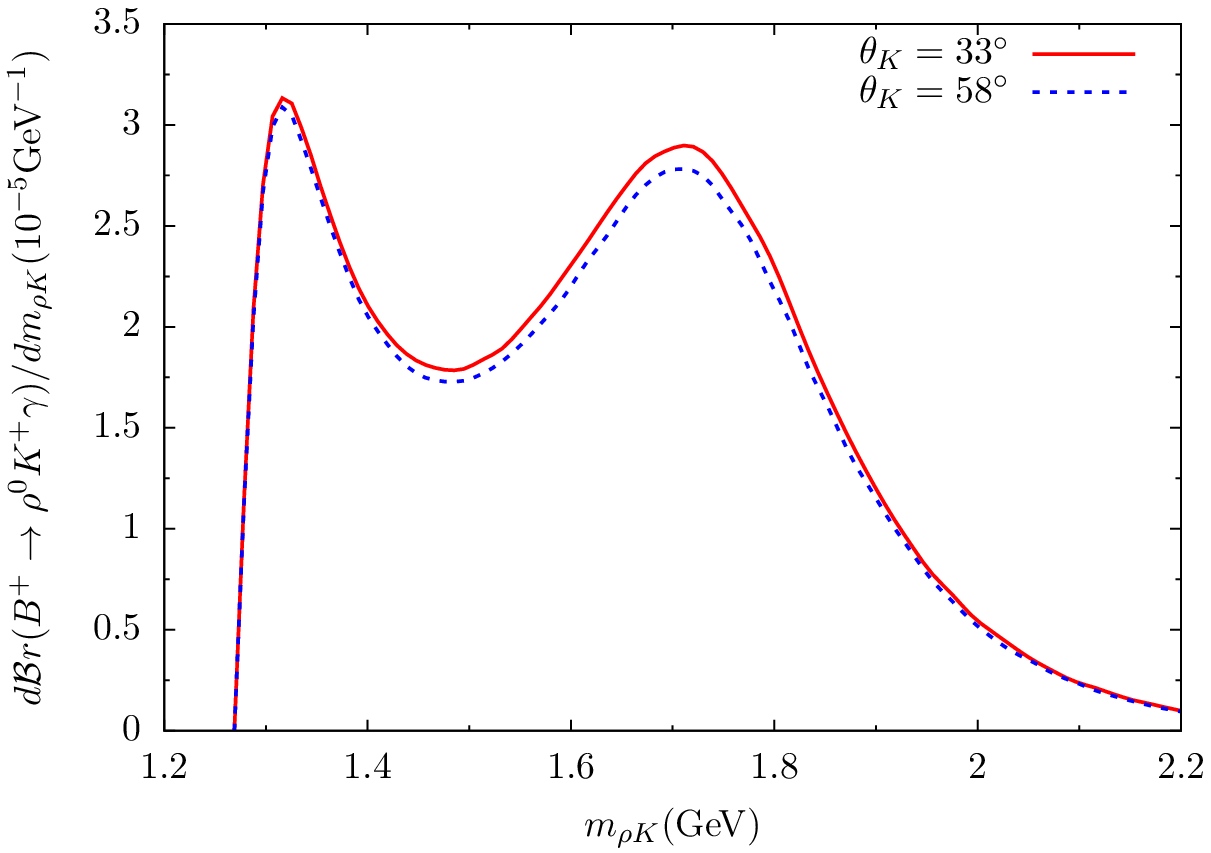}}
\subfigure[$B^0\to\rho^- K^+\gamma$]{
\includegraphics[width=7.5cm]{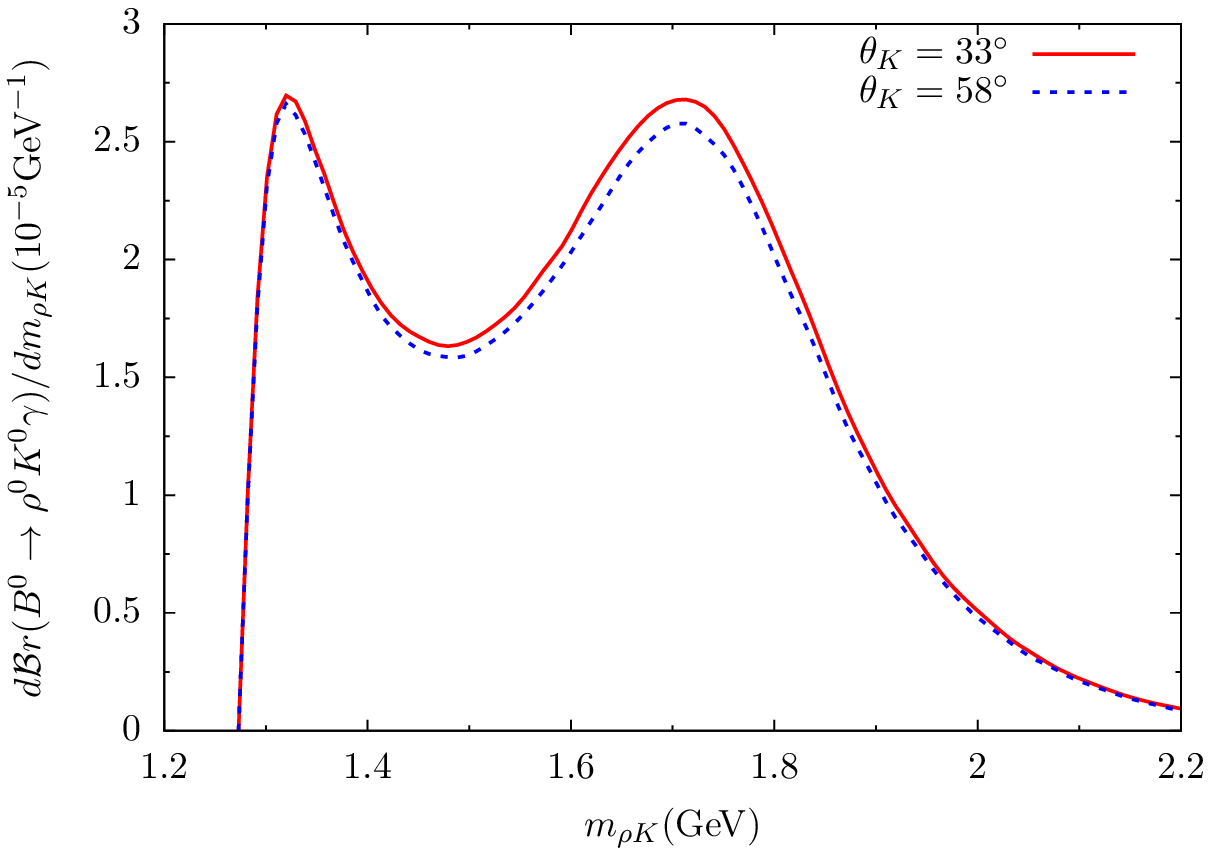}}
\caption{\label{spe2}Predicted $B\to\rho K\gamma$
decay spectra in the $\rho K$ invariant masses with the $K_1$ mixing
angle $\theta_K=33^\circ (c=2.0)$  and $58^\circ (c=1.8)$, respectively. }
\end{figure*}

At last, Fig.~\ref{spe1} shows our predictions for the $\phi K$ mass
distributions in the $B \to \phi K\gamma$ decays, in which the points with
error bars represent the Belle data~\cite{Sahoo:2011zd}
normalized to the central values
$\mathcal{B}(B^+ \to \phi K^+ \gamma)=2.48\times 10^{-6}$ and
$\mathcal{B}(B^0 \to  \phi K^0 \gamma)=2.74\times 10^{-6}$.
The comparison indicates the consistency with the Belle measurements:
both the predicted and
observed $B^+\to\phi K^+ \gamma$ spectra reach the maximum at around
$m_{\phi K}\sim 1.8$ GeV after a leap from the threshold. The peak
position accords the qualitative argument in the PQCD approach~\cite{Chen:2002th}
that the dominant nonresonant contributions to three-body $B$ meson decays
arise from the region with the invariant mass about $O(\bar{\Lambda}m_B)$,
$\bar{\Lambda}=m_B-m_b$ being the $B$ meson and $b$ quark mass difference.
The predicted $B\to \rho K\gamma$ decay spectra, presented in
Fig.~\ref{spe2}, exhibit two peaks around
the $K_1(1270)$ and $K^*(1680)$ masses as expected. Our predictions
for the above decay spectra can be confronted with future data.

\section{summary}\label{sec-5}

In this paper we have explored the three-body radiative decays $B\to PV\gamma$
in the PQCD framework, concentrating on the $B\to \phi(\rho) K\gamma$ modes.
The dominant contributions to three-body $B$ meson decays originate from the
regions corresponding to edges of Dalitz plots, where two final state
mesons are nearly collimated with each other. The $PV$ DAs have been
introduced to absorb the infrared dynamics in the meson pair, so that
a three-body decay amplitude can be factorized, similar to the two-body case,
into the convolution of the $PV$ DAs and hard kernels. We have extracted the
dependence of the $PV$ DAs on the meson momentum fraction through their
normalizations to the time-like form factors, and proposed appropriate
parametrizations for the nonsonant and resonant contributions. For the
$B\to\phi K\gamma$ decays, the nonresonant contributions dominate, and
the prominent feature of the decay spectra is the enhancement near the threshold.
For the $B\to\rho K\gamma$ decays, we have adopted the Breit-Wigner model
with a tunable parameter to characterize the relative strength between the
$K_{1}(1270)$ and $K^*(1680)$ states.

Fitting the PQCD factorization formulas to the branching-ratio data,
we have fixed the free parameters in the $PV$ DAs, which were then
employed to predict the direct $CP$ asymmetries,
the decay spectra and the photon polarization parameter
of the $B\to \phi(\rho) K\gamma$ modes. The $O_{7\gamma}$,
$O_{8g}$, and $O_{2}$ operators, and the annihilation contributions have
been taken into account, so this work is more complete than in the
literature~\cite{Chen:2004az}, where only the emission diagrams from
$O_{7\gamma}$ were considered. It has been shown that our results
are in good agreement with all the existing data. More precise data from
future experiments will help testing our predictions, including other
minor resonant contributions which have been ignored here, and improving the
application of the PQCD formalism to more three-body $B$ meson decays.

The analysis of the $B\to K^*\pi\gamma$ decays is similar, but requires
the inclusion of all the $K_1(1270)$, $K_1(1400)$, $K^*(1410)$, and
$K^*(1680)$ intermediate resonances~\cite{Sanchez:2015pxu}. Five
parameters are then needed to describe the interference among the resonances
with the same spin parity, three of them accounting for the
magnitudes and two for the phases. The present data are not sufficient
to determine these parameters, so we will leave the $B\to K^*\pi\gamma$ modes to a
future investigation.

\begin{acknowledgments}
We thank W.-F. Wang and W. Wang for useful discussions.
This work was supported in part by
the Ministry of Science and Technology of R.O.C. under Grant No.
MOST-104-2112-M-001-037-MY3, and by National Natural Science Foundation of China under Grants
No. 11575151, No. 11375208, No. 11521505, No. 11621131001, and No. 11235005.

\end{acknowledgments}

\appendix

\section{The factorization formalism}

The effective Hamiltonian relevant to the $b\to s$ transition
is given by~\cite{Buchalla:1995vs}
\begin{eqnarray}
H_{eff}&=&\frac{G_F}{\sqrt{2}}\big [\sum_{q=u,c}V_{qb}V_{qs}^*
\{C_1(\mu)O_1^{(q)}(\mu)+C_2(\mu)O_2^{(q)}(\mu)\}\nonumber \\
&&-V_{tb}V_{ts}^*
\sum_{i=3\sim 8g}C_i(\mu)O_i(\mu) \big ]+\text{H.c.},
\end{eqnarray}
with the Wilson coefficients $C$ and the local operators
\begin{eqnarray}
O_1^{(q)}&=&(\bar{s}_iq_j)_{V-A}(\bar{q}_jb_i)_{V-A},
\quad O_2^{(q)}=(\bar{s}_iq_i)_{V-A}(\bar{q}_jb_j)_{V-A},
\quad O_3=(\bar{s}_ib_i)_{V-A}\sum_q(\bar{q}_jq_j)_{V-A}, \nonumber\\
O_4&=&(\bar{s}_ib_j)_{V-A}\sum_q(\bar{q}_jq_i)_{V-A},
\quad O_5=(\bar{s}_ib_i)_{V-A}\sum_q(\bar{q}_jq_j)_{V+A},
\quad O_6=(\bar{s}_ib_j)_{V-A}\sum_q(\bar{q}_jq_i)_{V+A}, \nonumber \\
O_{7\gamma}&=&\frac{e}{8\pi^2}m_b\bar{s}_i\sigma^{\mu\nu}(1+\gamma_5)b_iF_{\mu\nu},
\quad O_{8g}=\frac{g}{8\pi^2}m_b\bar{s}_i\sigma^{\mu\nu}(1+\gamma_5)T_{ij}^ab_jG_{\mu\nu}^a,
\end{eqnarray}
where the terms associated with the strange quark mass
in the $O_{7\gamma}$ and $O_{8g}$ operators have been dropped.

\begin{figure*}
\begin{tabular}{cc}
\includegraphics[width=50mm] {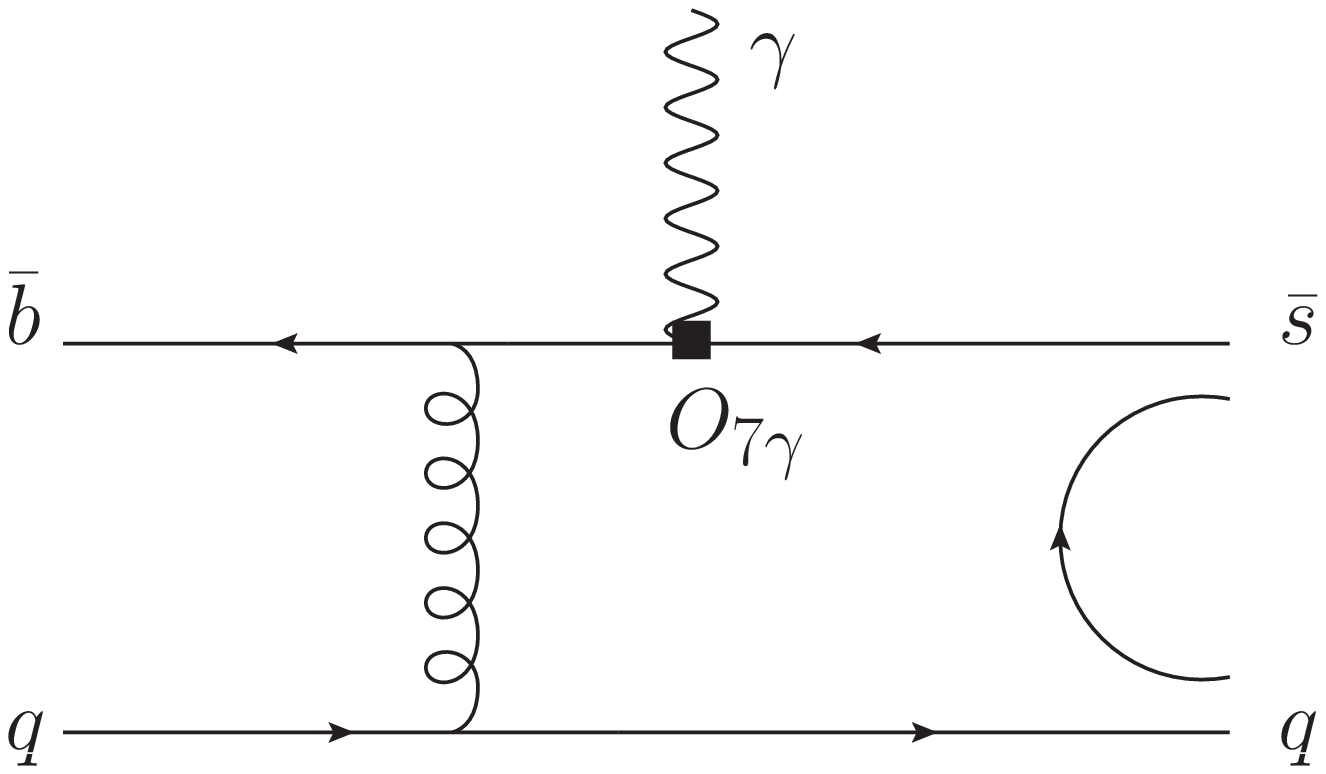} & \includegraphics[width=50mm] {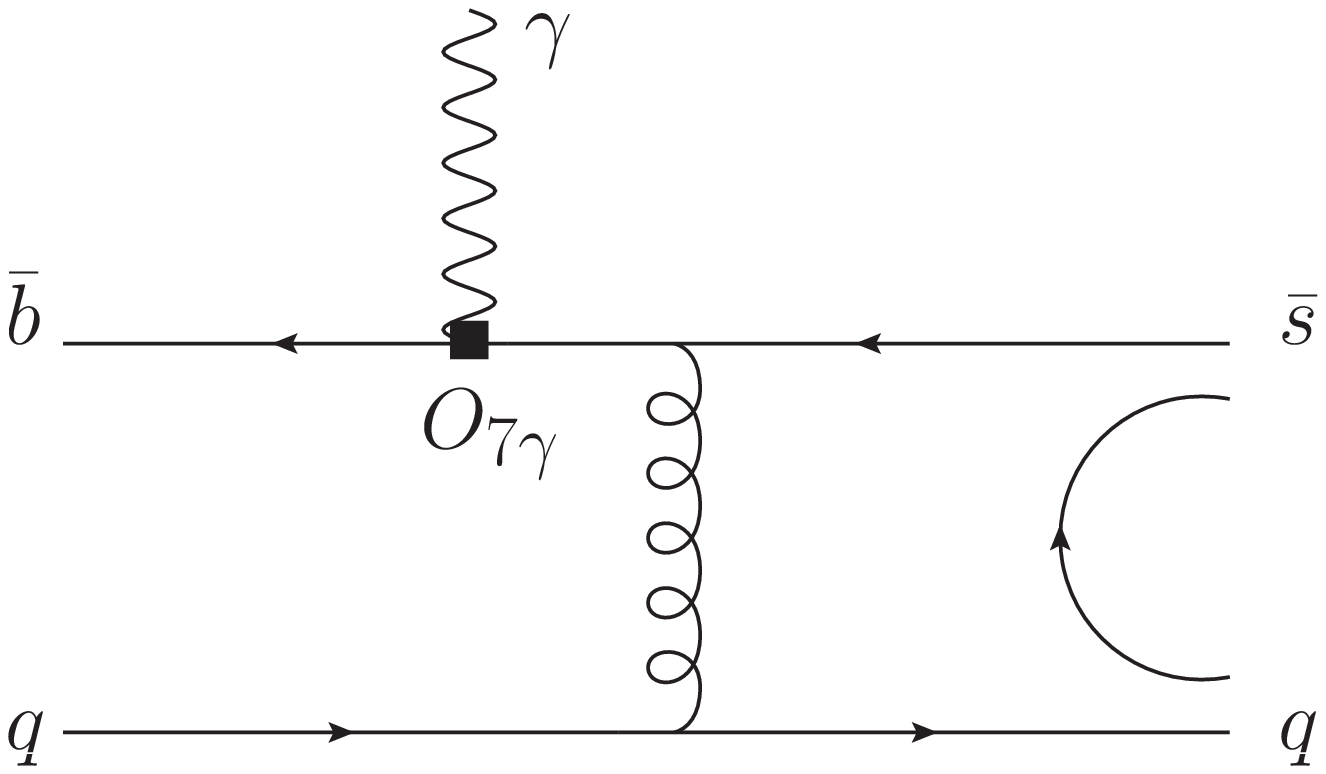}
\end{tabular}
\caption{\label{fo7}Feynman diagrams from the operator $O_{7\gamma}$.}
\end{figure*}

The dominant contributions to the three-body radiative decays $B\to PV\gamma$
comes from $O_{7\gamma}$, whose diagrams are displayed
in Fig.~\ref{fo7} with the photon being emitted from the operator.
The factorization formulas for the emissions of the right-handed and left-handed
photons are written as
\begin{eqnarray}
\mathcal{M}_{7\gamma}^R&=&-4C_F\frac{e}{\pi}m_bm_B^4
\int_0^1 dx_1dz\int_0^\infty b_1db_1b_2db_2\phi_B(x_1,b_1)(1-\eta)\Big\{[(1+z)\phi_t \\
&&+\sqrt{\eta}(1-2z)(\phi_v+\phi_a)] E_{e}(t_a)h_a(x_1,z,b_1,b_2)
+\sqrt{\eta}(\phi_v+\phi_a)E_{e}(t_a')h_a'(x_1,z,b_1,b_2)\Big\}, \nonumber \\
\mathcal{M}_{7\gamma}^L&=&0,
\end{eqnarray}
respectively, where the left-helicity amplitude $\mathcal{M}_{7\gamma}^L$ vanishes
because of the neglect of the strange quark mass.

\begin{figure*}
\begin{tabular}{cc}
\includegraphics[width=50mm] {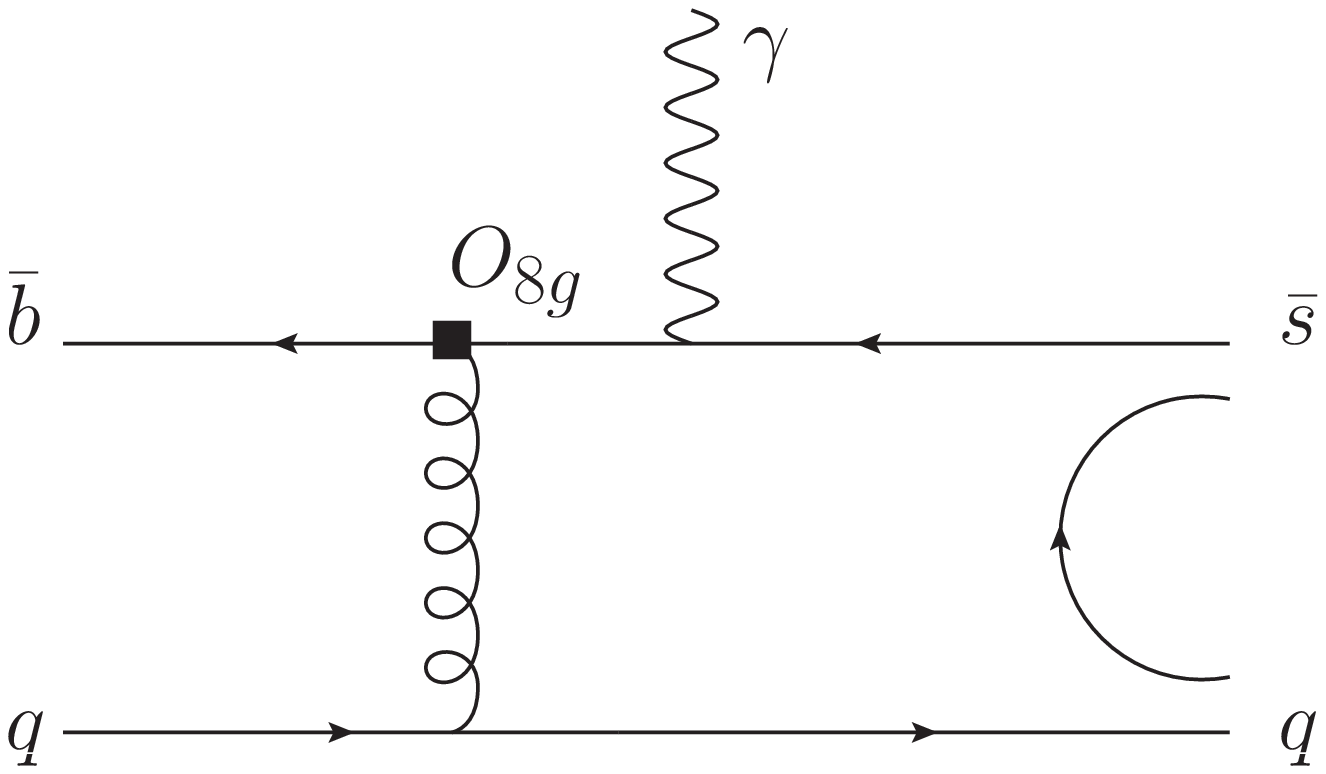} & \includegraphics[width=50mm, height=30mm] {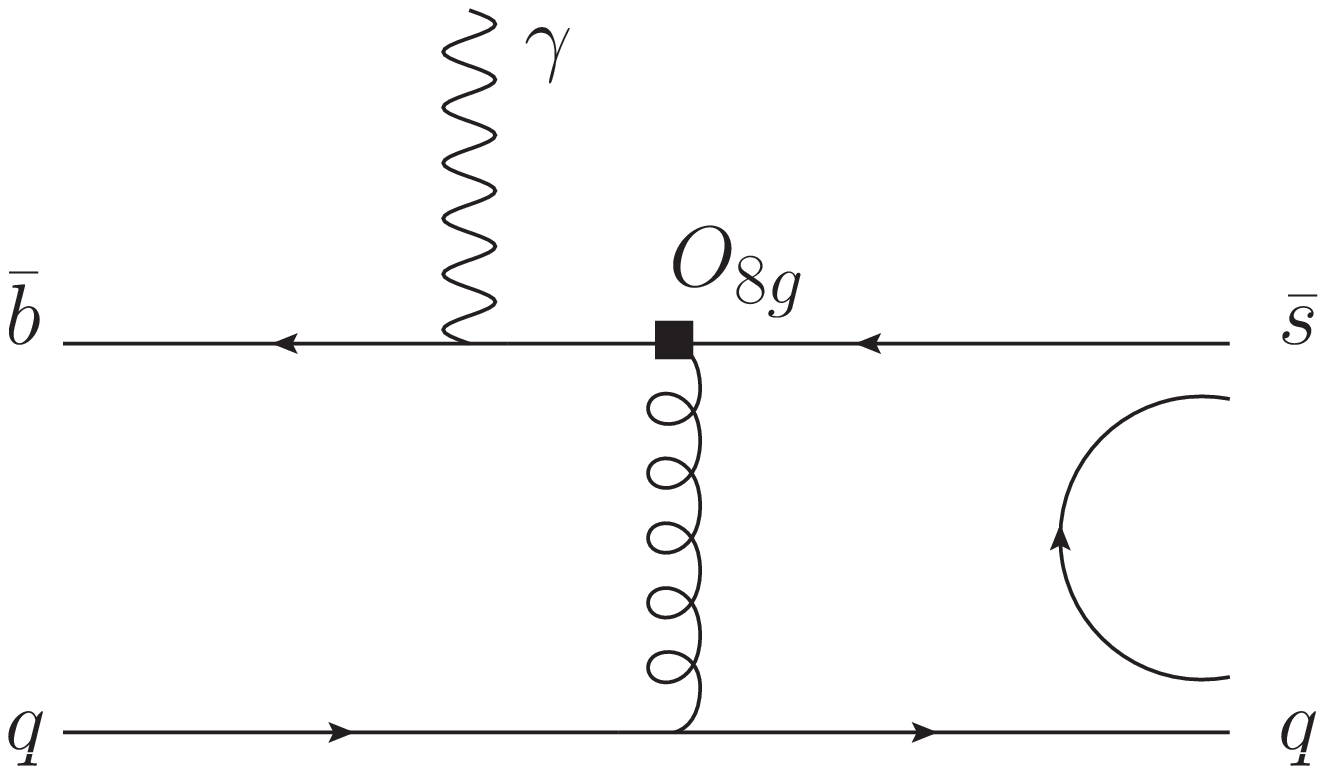}\\
 \\ \\
\includegraphics[width=50mm] {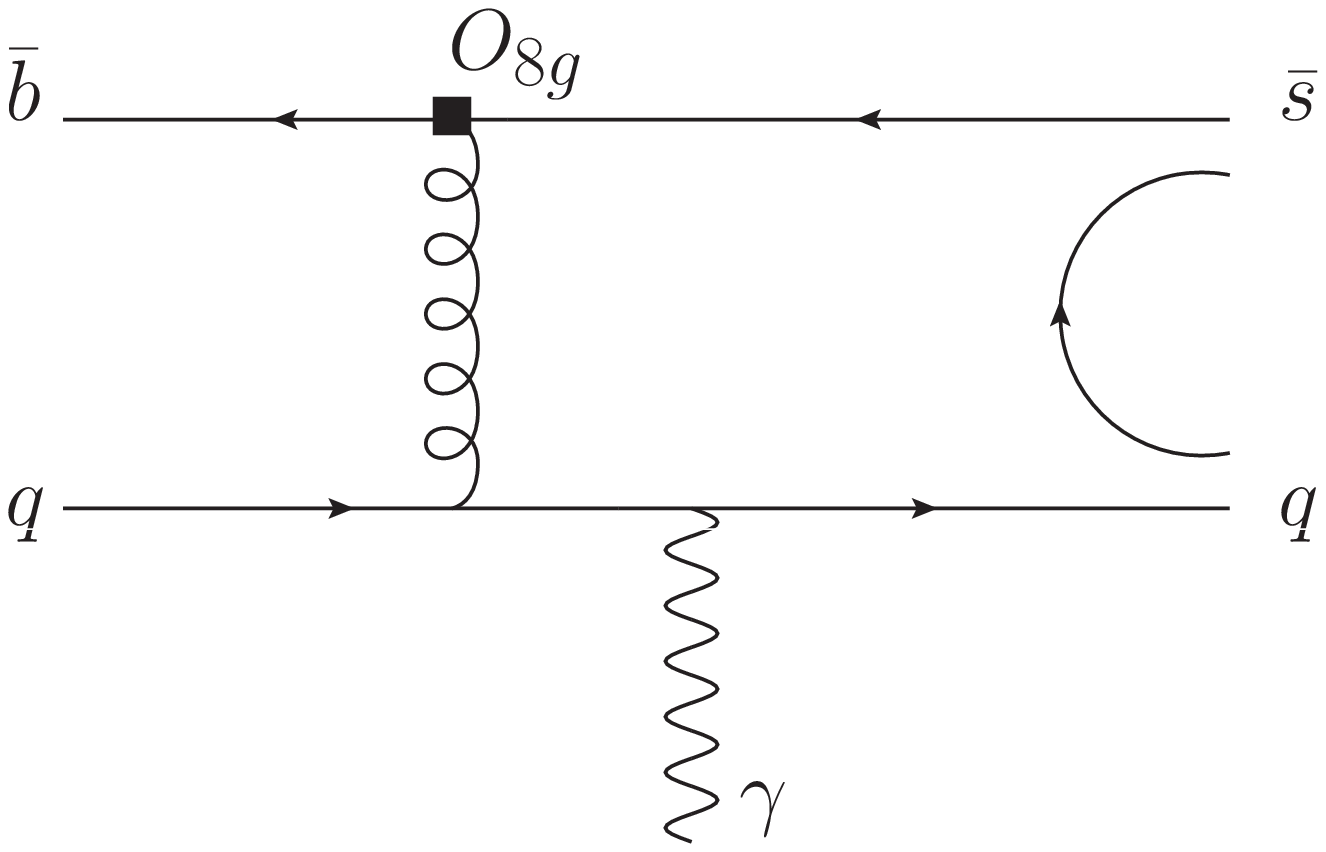} & \includegraphics[width=50mm, height=30mm] {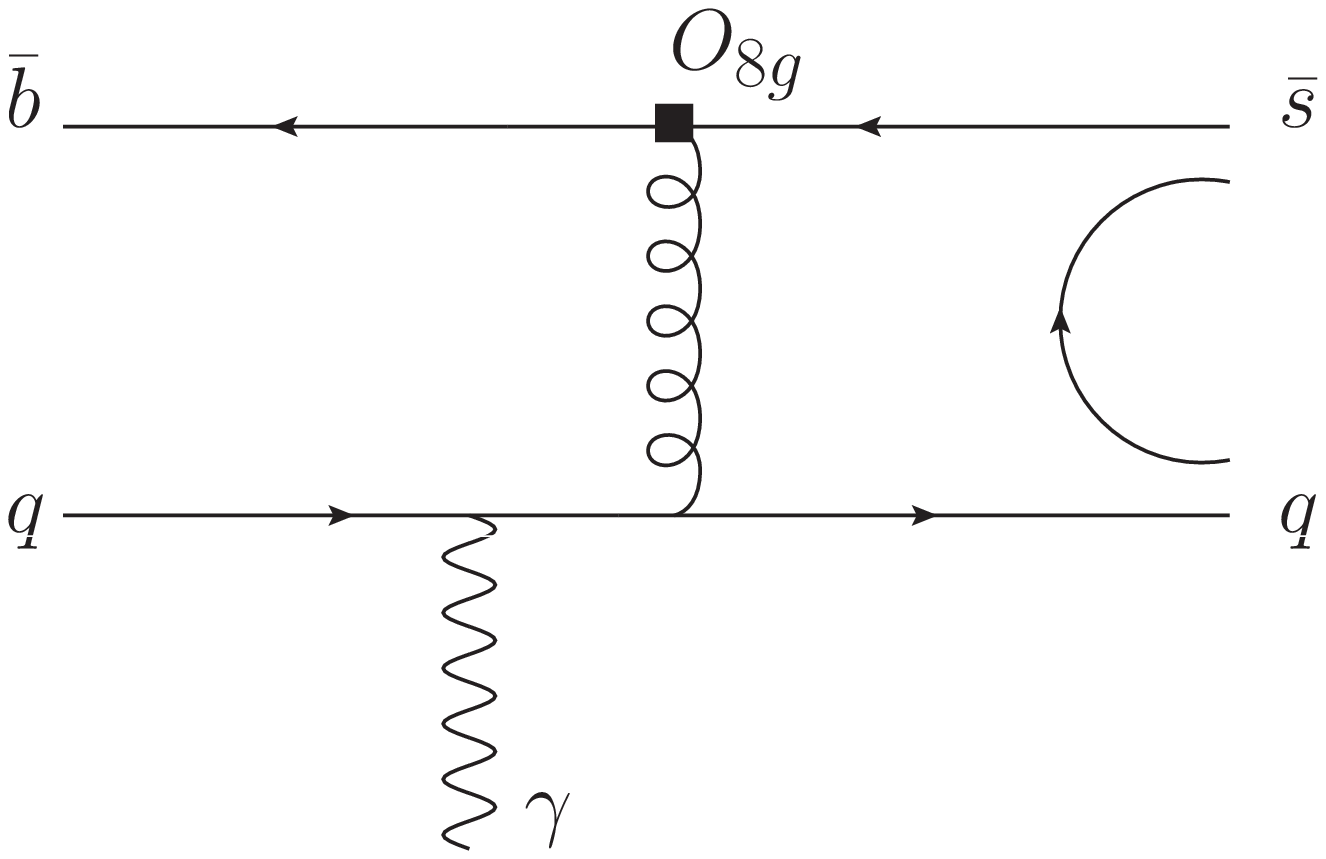}
\end{tabular}
\caption{\label{fo8}Feynman diagrams from the operator $O_{8g}$.}
\end{figure*}

The diagrams associated with the operator $O_{8g}$ are depicted in Fig.~\ref{fo8},
where the hard gluon from the operator $O_{8g}$ kicks the soft
spectator, making it an energetic collinear quark, and the photon is emitted via
the bremsstrahlung. The amplitudes are written as
\begin{eqnarray}
\mathcal{M}_{8g}^{R(a)}&=&\frac{2 C_F}{\pi}m_bm_B^4e
\int_0^1 dx_1dz\int_0^\infty b_1db_1b_2db_2\phi_B(x_1,b_1)
\Big\{ [(x_1-2z)\phi_t+3\sqrt{\eta}z(\phi_v+\phi_a)]Q_sE_e(t_b)\nonumber\\
&& \times h_b(x_1,z,b_1,b_2) -[(1-\eta+x_1)
(x_1\phi_t+\sqrt{\eta}z(\phi_v+\phi_a))]Q_bE_e(t_b')h_b'(x_1,z,b_1,b_2)\Big\},\\
\mathcal{M}_{8g}^{L(a)}&=&\frac{2 C_F}{\pi}m_bm_B^4Q_se
\int_0^1 dx_1dz\int_0^\infty b_1db_1b_2db_2\phi_B(x_1,b_1)\Big\{(1-z)(\eta(2x_1-z)\phi_t \nonumber \\
&&+3x_1\sqrt{\eta}(\phi_v-\phi_a))\Big\}E_e(t_b)h_b(x_1,z,b_1,b_2),
\end{eqnarray}
for the first two diagrams, where $Q_{b(s)}$ labels the charge of the
$b$ ($s$) quark in units of the electron charge $e$, and as
\begin{eqnarray}
\mathcal{M}_{8g}^{R(b)}(Q_q)&=&\frac{2 C_F}{\pi}m_bm_B^4Q_qe
\int_0^1 dx_1dz\int_0^\infty b_1db_1b_2db_2\phi_B(x_1,b_1)\Big\{ -(1-\eta)[(2+z-2\eta-x_1)\phi_t\nonumber\\
&& +3\sqrt{\eta}z(\phi_v+\phi_a)] E_e(t_c)h_c(x_1,z,b_1,b_2)
+[(1-\eta)(x_1\phi_t+\sqrt{\eta}(x_1-z)(\phi_v+\phi_a))]\nonumber \\
&&\times E_e(t_c')h_c'(x_1,z,b_1,b_2)\Big\},\\
\mathcal{M}_{8g}^{L(b)}(Q_q)&=&\frac{2 C_F}{\pi}m_bm_B^4Q_qe
\int_0^1 dx_1dz\int_0^\infty b_1db_1b_2db_2\phi_B(x_1,b_1)
\Big\{[-z\eta(1+2z-2x_1-\eta)\phi_t\nonumber\\
&&+3\sqrt{\eta}(1-\eta)(z-x_1)(\phi_v-\phi_a)] E_e(t_c)h_c(x_1,z,b_1,b_2)
+[(1-\eta)(\eta(x_1-z)\phi_t \nonumber \\
&&-\sqrt{\eta}x_1(\phi_v-\phi_a))]E_e(t_c')h_c'(x_1,z,b_1,b_2)\Big\},
\end{eqnarray}
for the last two diagrams.

Next we consider the quark loop corrections from the operators $O_i$.
The operator $O_1$ does not contribute due to the color mismatch,
and the $O_{3\sim6}$ insertions are small compared to the $O_2$ insertion.
Hence, we consider only the $O_2$ contributions, in which
the photon is emitted either by an external quark or from the quark loop.

\begin{figure*}
\begin{tabular}{cc}
\includegraphics[width=50mm] {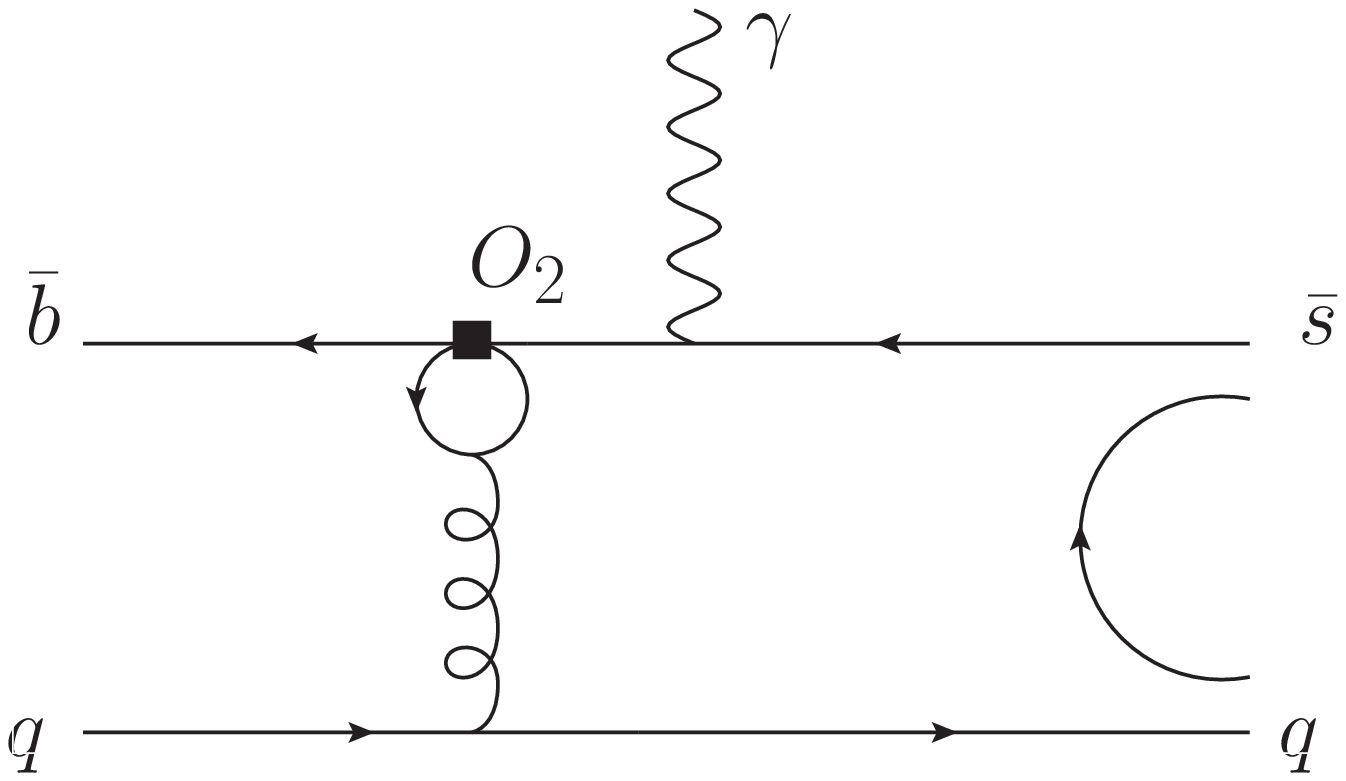} & \includegraphics[width=50mm] {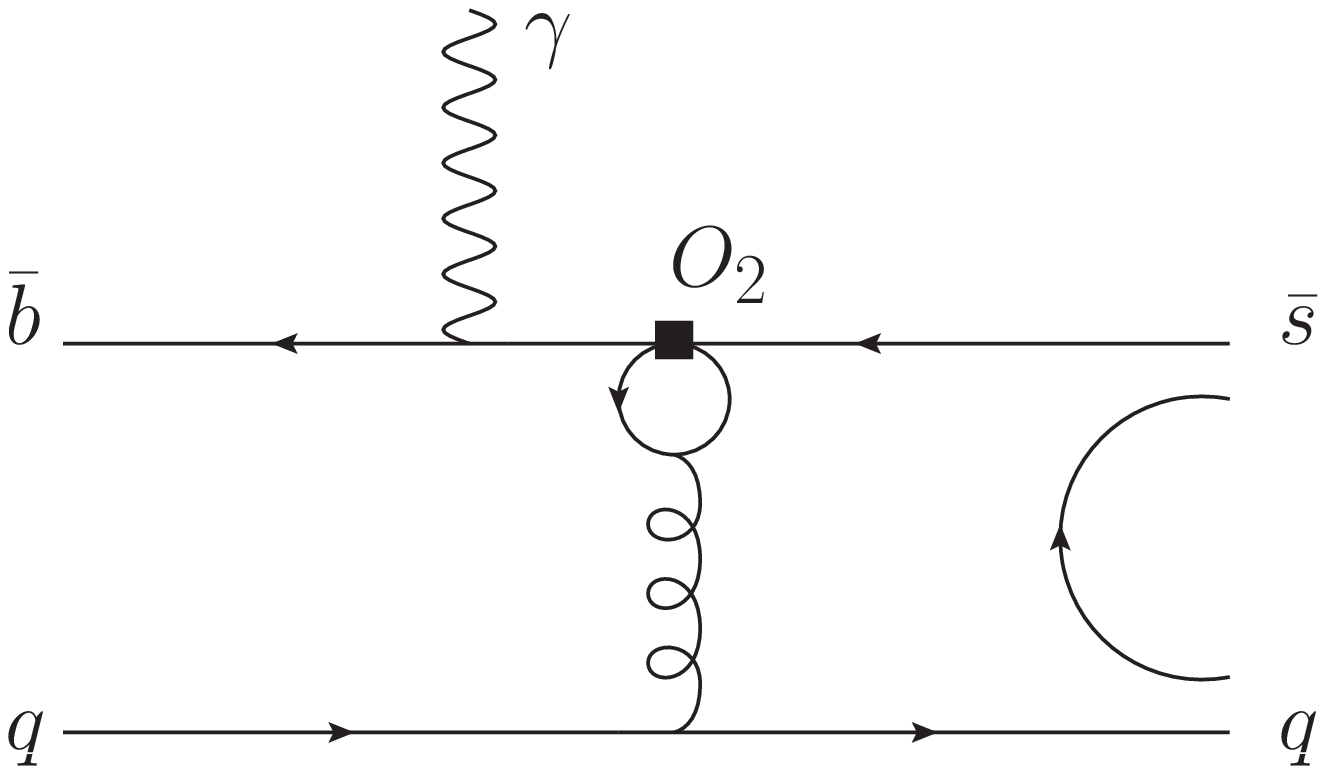}\\
 \\ \\
\includegraphics[width=50mm] {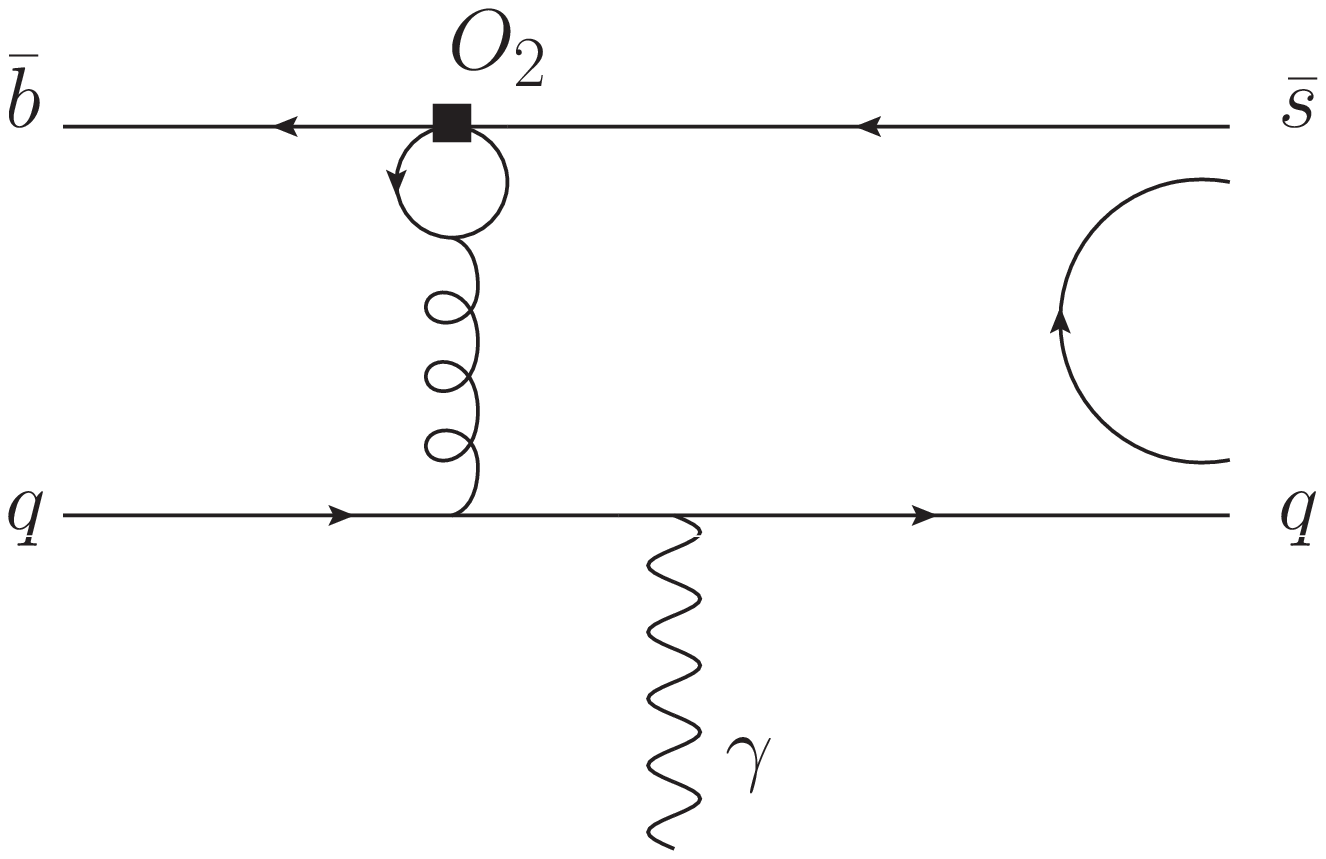} & \includegraphics[width=50mm] {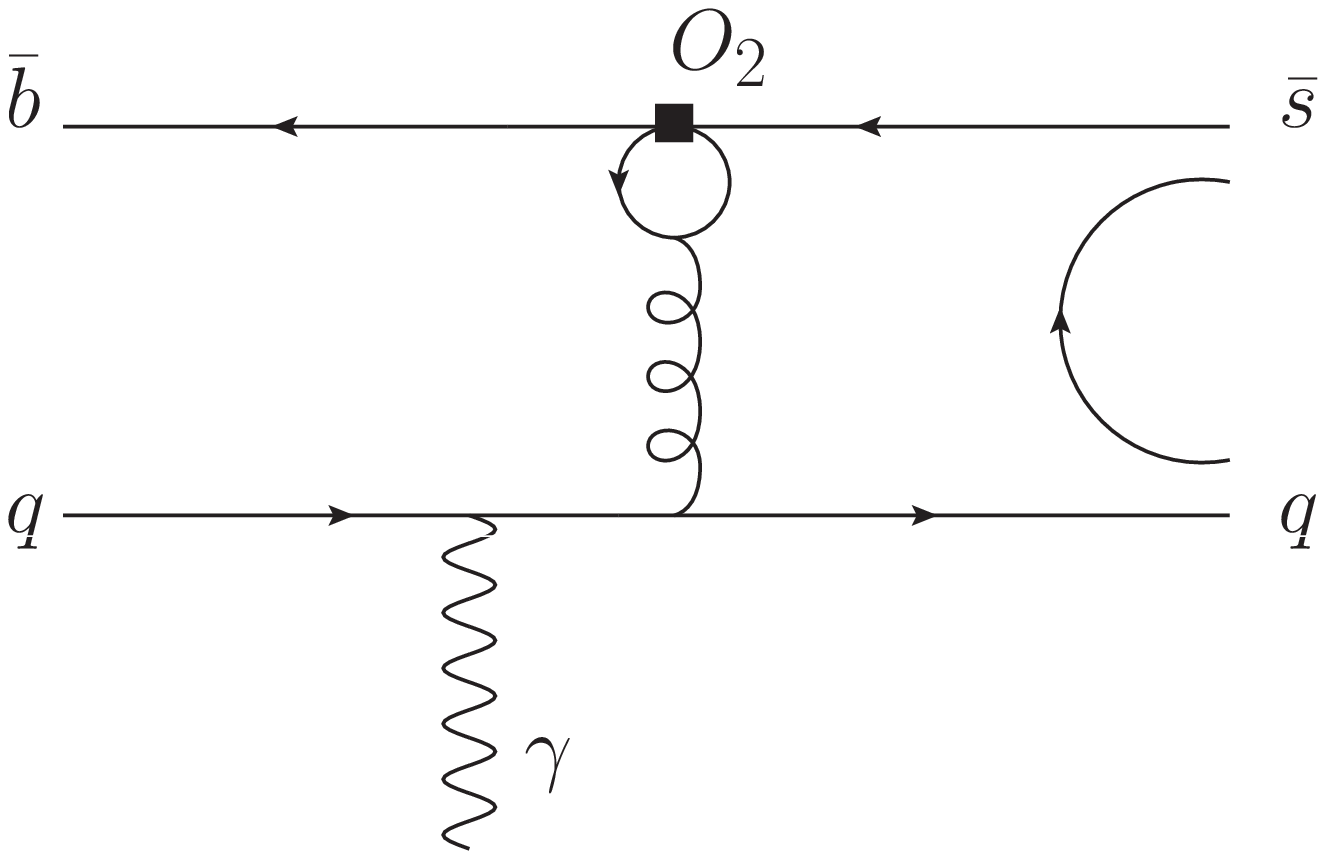}
\end{tabular}
\caption{\label{fo2}Quark-loop diagrams from the operator $O_{2}$ with a photon
being emitted by an external quark.}
\end{figure*}

The diagrams with the photon being emitted by an external quark
are shown in Fig.~\ref{fo2}. The effective vertex $\bar{b}\to \bar{s}g$
resulting from the loop integration in the $\overline{MS}$ scheme
is given by~\cite{Bander:1979px}
\begin{eqnarray}
I^\nu&=&-\frac{g}{8\pi^2}\left[G(m_i^2,k^2,\mu)-\frac{2}{3}\right]
\bar{b}T^a(k^2\gamma^\nu-k^\nu\slashed{k})(1-\gamma_5)s, \nonumber \\
G(m_i^2,k^2,\mu)&=&-\int_0^1dx4x(1-x)\text{log}
\left[\frac{m_i^2-x(1-x)k^2-i\epsilon}{\mu^2}\right],
\end{eqnarray}
where $k$ is the virtual gluon momentum and
$m_i$, $i=u,c$, are the masses of the quarks in the loop.
The amplitudes are written as
\begin{eqnarray}
\mathcal{M}_{1i}^{R(a)}&=&\frac{C_F}{\pi}m_B^5Q_se
\int_0^1 dx_1dz\int_0^\infty b_1db_1b_2db_2
\phi_B(x_1,b_1)[G(m_i^2,-x_1zm_B^2,t_b)-\frac{2}{3}] \nonumber\\
&&\times z(3x_1\phi_t+\sqrt{\eta}(z-2x_1)(\phi_v+\phi_a))
E_e(t_b)h_b(x_1,z,b_1,b_2),\\
\mathcal{M}_{1i}^{L(a)}&=&\frac{C_F}{\pi}m_B^5e
\int_0^1 dx_1dz\int_0^\infty b_1db_1b_2db_2
\phi_B(x_1,b_1)\Big\{[G(m_i^2,-x_1zm_B^2,t_b)-\frac{2}{3}]x_1(1-z) \nonumber\\
&&\times (3\eta z\phi_t+\sqrt{\eta}(2z-x_1)(\phi_v-\phi_a))
Q_sE_e(t_b)h_b(x_1,z,b_1,b_2)-[G(m_i^2,-x_1zm_B^2,t_b') \nonumber \\
&&-\frac{2}{3}](z(1-\eta+x_1)(\eta z\phi_t-x_1
\sqrt{\eta}(\phi_v-\phi_a)))Q_bE_e(t_b')h_b'(x_1,z,b_1,b_2)\Big\},
\end{eqnarray}
for the first two diagrams, and as
\begin{eqnarray}
\mathcal{M}_{1i}^{R(b)}(Q_q)&=&-\frac{C_F}{\pi}m_B^5Q_qe
\int_0^1 dx_1dz\int_0^\infty b_1db_1b_2db_2\phi_B(x_1,b_1)
\Big\{[G(m_i^2,(z-x_1)(1-\eta)m_B^2,t_c) \nonumber\\
&&-\frac{2}{3}](1-\eta)[3(z-x_1)(1-\eta)\phi_t+z
\sqrt{\eta}(1+2z-\eta-2x_1)(\phi_v+\phi_a)]E_e(t_c)h_c(x_1,z,b_1,b_2) \nonumber \\
&&-[G(m_i^2,(z-x_1)(1-\eta)m_B^2,t_c')-\frac{2}{3}]
((1-\eta)^2(x_1\phi_t+\sqrt{\eta}(x_1-z)(\phi_v+\phi_a))) \nonumber \\
&&\times E_e(t_c')h_c'(x_1,z,b_1,b_2)\Big\},\\
\mathcal{M}_{1i}^{L(b)}(Q_q)&=&-\frac{C_F}{\pi}m_B^5Q_qe
\int_0^1 dx_1dz\int_0^\infty b_1db_1b_2db_2\phi_B(x_1,b_1)
\Big\{[G(m_i^2,(z-x_1)(1-\eta)m_B^2,t_c) \nonumber\\
&&-\frac{2}{3}] (1-\eta)(z-x_1)[3\eta z\phi_t-\sqrt{\eta}(2+z-2\eta-x_1)
(\phi_v-\phi_a)]E_e(t_c)h_c(x_1,z,b_1,b_2) \nonumber \\
&&+[G(m_i^2,(z-x_1)(1-\eta)m_B^2,t_c')-\frac{2}{3}]
[(1-\eta)(x_1-z)(\eta(x_1-z)\phi_t -\sqrt{\eta}x_1(\phi_v-\phi_a))] \nonumber \\
&&\times E_e(t_c')h_c'(x_1,z,b_1,b_2)\Big\},
\end{eqnarray}
for the last two diagrams.

\begin{figure*}
\begin{tabular}{cc}
\includegraphics[width=60mm] {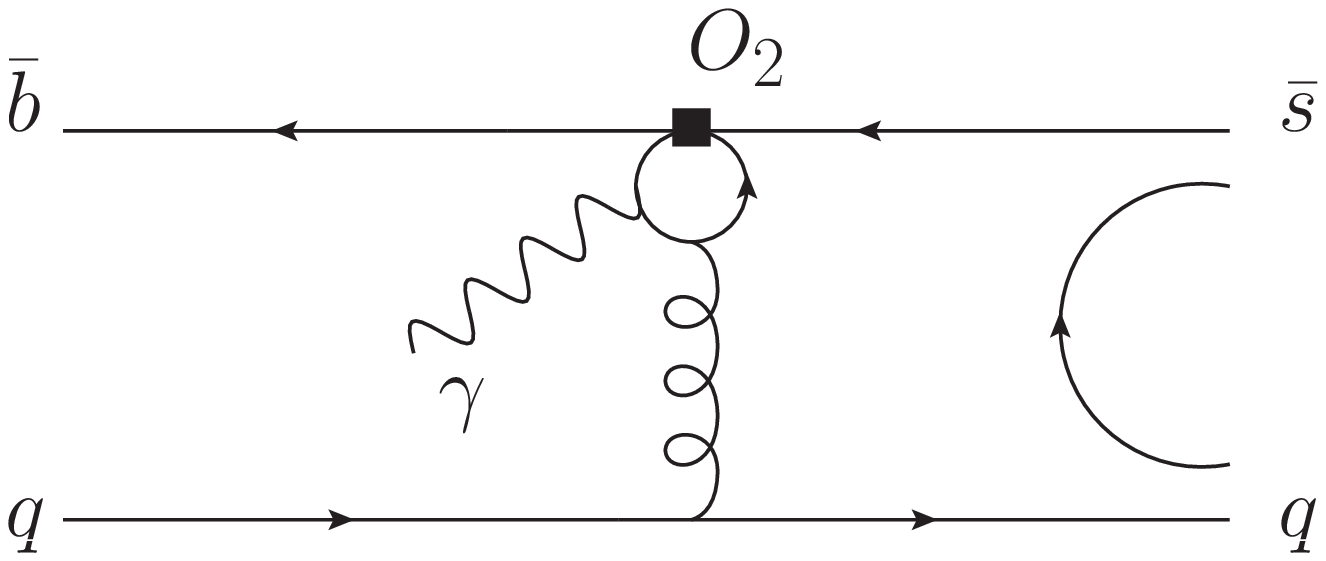} & \includegraphics[width=60mm] {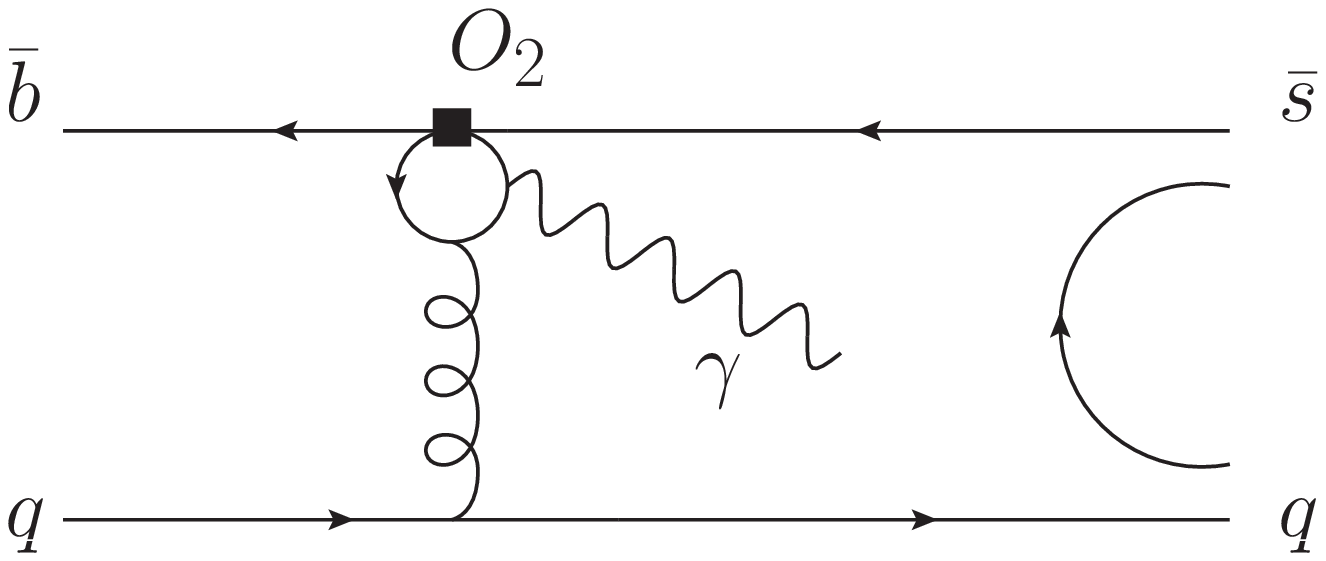}
\end{tabular}
\caption{\label{fo72}Quark-loop diagrams from the operator $O_{2}$ with the photon
being emitted from the quark loop.}
\end{figure*}

In the case where the photon is emitted from the quark loop, as displayed
in Fig.~\ref{fo72}, the sum of the effective vertex $\bar{b}\to\bar{s}\gamma g^*$
produces~\cite{Liu:1990yb,Simma:1990nr}
\begin{eqnarray}
I=\bar{b}\gamma^\rho\frac{(1-\gamma_5)}{2}T^asI_{\mu\nu\rho}A^{\mu}A^{a\nu},
\end{eqnarray}
where
\begin{eqnarray}
I_{\mu\nu\rho}&=&A_4[(q\cdot k)\epsilon_{\mu\nu\rho\sigma}
(q-k)^\sigma+\epsilon_{\nu\rho\sigma\tau}q^\sigma k^\tau k_\mu
-\epsilon_{\mu\nu\sigma\tau}q^\sigma k^\tau q_\nu] \nonumber \\
&&+A_5[\epsilon_{\mu\rho\sigma\tau}q^\sigma k^\tau k_\nu-k^2\epsilon_{\mu\nu\rho\sigma}q^\tau],
\end{eqnarray}
with
\begin{eqnarray}
A_4&=&-\frac{4ieg}{3\pi^2}\int_0^1dx\int_0^{1-x}dy\frac{xy}{x(1-x)k^2+2xyq\cdot k-m_i^2},\\
A_5&=&\frac{4ieg}{3\pi^2}\int_0^1dx\int_0^{1-x}dy\frac{x(1-x)}{x(1-x)k^2+2xyq\cdot k-m_i^2},
\end{eqnarray}
$q$ ($k$) being the photon (gluon) momentum.
The amplitudes for the two diagrams are expressed as
\begin{eqnarray}
\mathcal{M}_{2i}^{R}&=&-\frac{8}{3}\frac{C_F}{\pi}m_B^5e
\int_0^1dx\int_0^{1-x}dy\int_0^1 dx_1dz\int_0^\infty b_1db_1
\phi_B(x_1,b_1)\alpha_s(t_d)e^{-S_B(t_d)}\nonumber\\
&&\times \frac{h_e'}{xyz(1-\eta)m_B^2-m_i^2} \times (1-\eta)z
\Big\{xy[(1-\eta+2x_1)\phi_t-\sqrt{\eta}(1-\eta-z+x_1)(\phi_v+\phi_a)]\nonumber \\
&&-x(1-x)[3x_1\phi_t+\sqrt{\eta}(z-2x_1)(\phi_v+\phi_a)]\Big\},\\
\mathcal{M}_{2i}^{L}&=&\frac{8}{3}\frac{C_F}{\pi}m_B^5e
\int_0^1dx\int_0^{1-x}dy\int_0^1 dx_1dz\int_0^\infty b_1db_1
\phi_B(x_1,b_1)\alpha_s(t_d)e^{-S_B(t_d)} \nonumber \\
&&\times \frac{h_e'}{xyz(1-\eta)m_B^2-m_i^2}\times (1-\eta)z
\Big\{xy[2\eta z\phi_t+(z-x_1)\sqrt{\eta}(\phi_v-\phi_a)]-x(1-x) \nonumber \\
&&\times[\eta z\phi_t-x_1\sqrt{\eta}(\phi_v-\phi_a)] \Big\},
\end{eqnarray}
in which the function $h_e'$ is defined by
\begin{eqnarray}
h_e'=K_0(\sqrt{x_1z}m_Bb_1)-[\theta(B^2)K_0(b_1\sqrt{B^2})
+\theta(-B^2)\frac{i\pi}{2}H_0^{(1)}(b_1\sqrt{|B^2|})],
\end{eqnarray}
with
\begin{eqnarray}
B^2=x_1zm_B^2-\frac{yz(1-\eta)}{1-x}m_B^2+\frac{m_i^2}{x(1-x)}.
\end{eqnarray}

\begin{figure*}
\begin{tabular}{cc}
\includegraphics[width=50mm] {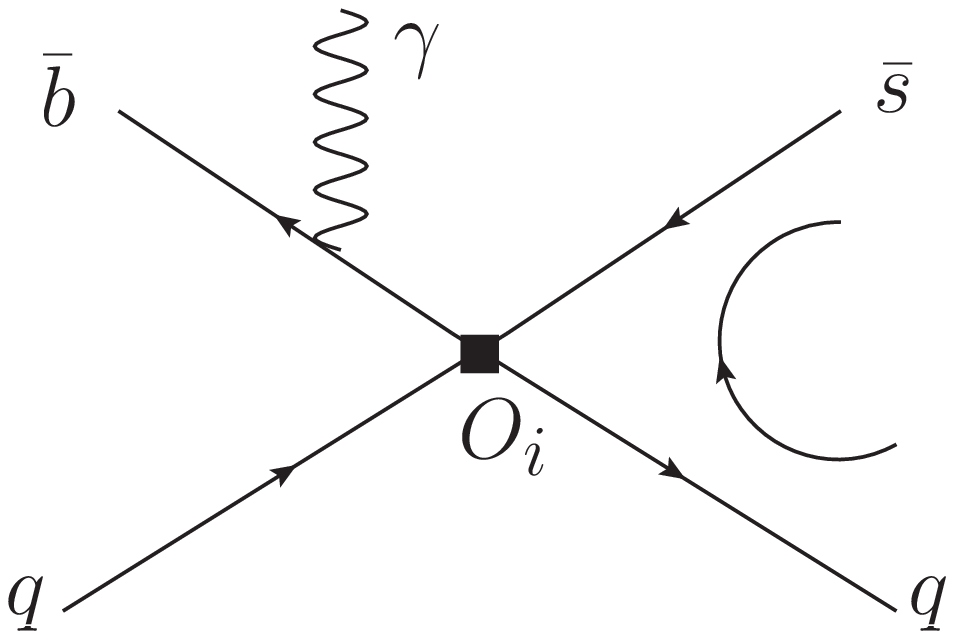} & \includegraphics[width=50mm] {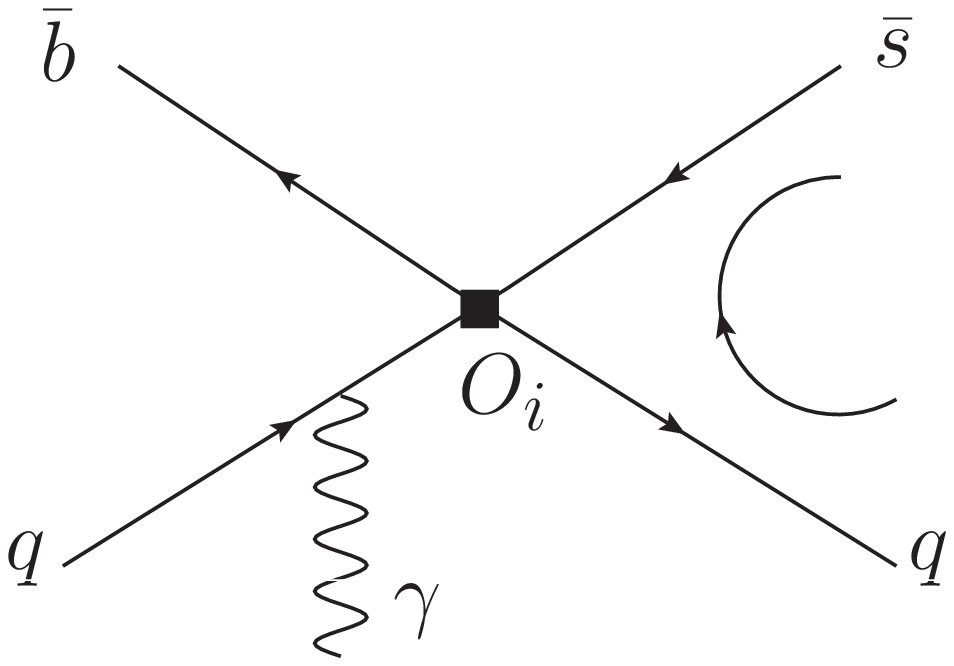}\\
 \\ \\
\includegraphics[width=50mm] {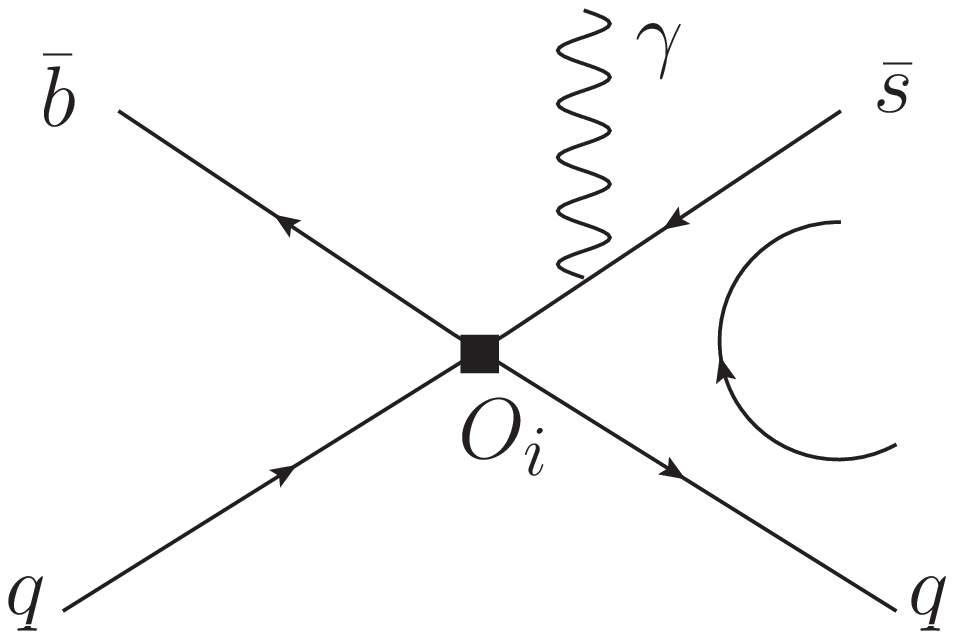} & \includegraphics[width=50mm] {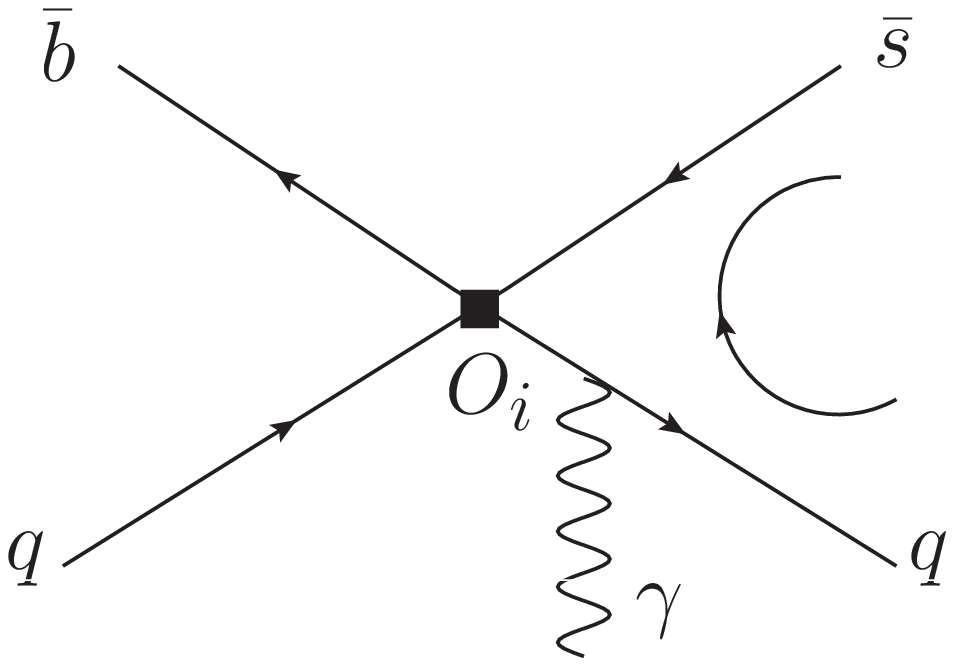}
\end{tabular}
\caption{\label{foan}Annihilation diagrams.}
\end{figure*}

The annihilation diagrams are exhibited in Fig.~\ref{foan}, to which
three types of operators contribute, the left-handed current between
$\bar{b}$ and $q$ quark and the left-handed current between the final
state quarks ($LL$); the left-handed current between $\bar{b}$ and $q$
quark and the right-handed current between the final state quarks ($LR$);
the $(S-P)(S+P)$ current from the Fierz transformation of the $(V-A)(V+A)$
operators (SP). Here we define the combinations of the Wilson coefficients:
\begin{eqnarray}
a_1&=&C_2+C_1/3, ~ a_4=C_4+C_3/3,~ a_6=C_6+C_5/3, \nonumber \\
a_8&=&C_8+C_7/3,~a_{10}=C_{10}+C_9/3.
\end{eqnarray}
The factorization formulas for the annihilation contributions are given by
\begin{eqnarray}
\mathcal{M}_{ann}^{R(a,LL)}(Q_q)&=&2\sqrt{6\eta}m_B^3Q_qe[(2\zeta-1)F_v+F_a]
\int_0^1dx\int_0^\infty b_1db_1\phi_B(x_1,b_1)\nonumber \\
&&\times E_a(t_e')K_0(\sqrt{x_1(1-\eta)}m_Bb_1)(1-\eta),\\
\mathcal{M}_{ann}^{L(a,LL)}(Q_q)&=&2\sqrt{6\eta}m_B^3Q_be[(2\zeta-1)F_v-F_a)
\int_0^1dx\int_0^\infty b_1db_1\phi_B(x_1,b_1) \\
&&\times \Big\{E_a(t_e)K_0(\sqrt{1-\eta+x_1}m_Bb_1)(1-\eta+x_1)
+E_a(t_e')K_0(\sqrt{x_1(1-\eta)}m_Bb_1)x_1\Big\},\nonumber \\
\mathcal{M}_{ann}^{R(a,LR)}(Q_q)&=&2\sqrt{6\eta}
m_B^3Q_qe[(2\zeta-1)F_v-F_a]\int_0^1dx\int_0^\infty b_1db_1\phi_B(x_1,b_1)\nonumber \\
&&\times E_a(t_e')K_0(\sqrt{x_1(1-\eta)}m_Bb_1)(1-\eta),\\
\mathcal{M}_{ann}^{L(a,LR)}(Q_q)&=&2\sqrt{6\eta}m_B^3Q_be[(2\zeta-1)F_v+F_a)
\int_0^1dx\int_0^\infty b_1db_1\phi_B(x_1,b_1) \\
&&\times \Big\{E_a(t_e)K_0(\sqrt{1-\eta+x_1}m_Bb_1)(1-\eta+x_1)
+E_a(t_e')K_0(\sqrt{x_1(1-\eta)}m_Bb_1)x_1\Big\}, \nonumber
\end{eqnarray}
\begin{eqnarray}
\mathcal{M}_{ann}^{R(b,LL)}(Q_q)&=&-\mathcal{M}_{ann}^{R,(b,LR)}(Q_q)\nonumber\\
&=&-2\sqrt{6\eta}ef_Bm_B^3\int_0^1dz\int_0^\infty b_2db_2
\Big\{Q_s(\phi_v+\phi_a)E_a'(t_f)\frac{i\pi}{2}H_0^{(1)}(\sqrt{1-z}m_Bb_2) \nonumber \\
&& -Q_qz(\phi_v+\phi_a)E_a'(t_f')\frac{i\pi}{2}H_0^{(1)}(\sqrt{z(1-\eta)}m_Bb_2)\Big\},\\
\mathcal{M}_{ann}^{L(b,LL)}(Q_q)&=&-\mathcal{M}_{ann}^{L,(b,LR)}(Q_q)\nonumber \\
&=&2\sqrt{6\eta}ef_Bm_B^3\int_0^1dz\int_0^\infty b_2db_2\Big\{Q_s(1-z)
(\phi_v-\phi_a)E_a'(t_f)\frac{i\pi}{2}H_0^{(1)}(\sqrt{1-z}m_Bb_2)\nonumber \\
&& -(1-\eta)Q_q(\phi_v-\phi_a)E_a'(t_f')\frac{i\pi}{2}H_0^{(1)}(\sqrt{z(1-\eta)}m_Bb_2)\Big\},
\end{eqnarray}
\begin{eqnarray}
\mathcal{M}_{ann}^{R(SP)}(Q_q)&=&4\sqrt{6}ef_Bm_B^3
\int_0^1dz\int_0^\infty b_2db_2\Big\{Q_s\phi_tE_a'(t_f)
\frac{i\pi}{2}H_0^{(1)}(\sqrt{1-z}m_Bb_2) \nonumber \\
&& +Q_q(1-\eta)\phi_tE_a'(t_f')\frac{i\pi}{2}H_0^{(1)}(\sqrt{z(1-\eta)}m_Bb_2)\Big\},\\
\mathcal{M}_{ann}^{L(SP)}(Q_q)&=&4\sqrt{6}ef_Bm_B^3
\int_0^1dz\int_0^\infty b_2db_2\Big\{Q_s\eta(1-z)\phi_tE_a'(t_f)
\frac{i\pi}{2}H_0^{(1)}(\sqrt{1-z}m_Bb_2) \nonumber \\
&&+Q_q\eta z\phi_tE_a'(t_f')\frac{i\pi}{2}H_0^{(1)}(\sqrt{z(1-\eta)}m_Bb_2)\Big\}.
\end{eqnarray}

Finally, we sum the squared amplitudes for the $B^+$ meson decays in the helicity basis,
$|\mathcal{A}(B^+)|^2=\sum_{i=R,L}|\mathcal{A}^i(B^+)|^2$,
deriving
\begin{eqnarray}
\mathcal{A}^i(B^+)&=&\frac{G_F}{\sqrt{2}}V_{ub}^*V_{us}
\Big\{C_2\left(\mathcal{M}_{1u}^{i(a)} +\mathcal{M}_{1u}^{i(b)}(Q_u)
+\mathcal{M}_{2u}^{i} \right) +a_1\left(\mathcal{M}_{ann}^{i(a,LL)}
(Q_u)+\mathcal{M}_{ann}^{i(b,LL)}(Q_u)\right)\Big\} \nonumber\\
&+&\frac{G_F}{\sqrt{2}}V_{cb}^*V_{cs}\Big\{C_2\left(\mathcal{M}_{1c}^{i(a)}
+\mathcal{M}_{1c}^{i(b)}(Q_u)+\mathcal{M}_{2c}^{i} \right)\Big\} \nonumber
\\ &-&\frac{G_F}{\sqrt{2}}V_{tb}^*V_{ts}\Big\{C_{7\gamma}\mathcal{M}_{7\gamma}^{i}
+C_{8g}\left(\mathcal{M}_{8g}^{i(a)}+\mathcal{M}_{8g}^{i(b)}(Q_u)\right) \nonumber \\
&+&(a_4+a_{10})\left(\mathcal{M}_{ann}^{i(a,LL)}(Q_u)+\mathcal{M}_{ann}^{i(b,LL)}(Q_u)\right)
+(a_6+a_8)\mathcal{M}_{ann}^{i(SP)}(Q_u)\Big\}. \label{e59}
\end{eqnarray}
We have the similar sum for the $B^0$ meson decay amplitudes with
\begin{eqnarray}
\mathcal{A}^i(B^0)&=&\frac{G_F}{\sqrt{2}}V_{ub}^*V_{us}
\Big\{C_2\left(\mathcal{M}_{1u}^{i(a)} +\mathcal{M}_{1u}^{i(b)}(Q_d)
+\mathcal{M}_{2u}^{i}\right)\Big\} \nonumber\\
&+&\frac{G_F}{\sqrt{2}}V_{cb}^*V_{cs}\Big\{C_2\left(\mathcal{M}_{1c}^{i(a)}
+\mathcal{M}_{1c}^{i(b)}(Q_d)+\mathcal{M}_{2c}^{i} \right)\Big\} \nonumber \\
&-&\frac{G_F}{\sqrt{2}}V_{tb}^*V_{ts}\Big\{C_{7\gamma}\mathcal{M}_{7\gamma}^{i}
+C_{8g}\left(\mathcal{M}_{8g}^{i(a)}+\mathcal{M}_{8g}^{i(b)}(Q_d)\right) \nonumber \\
&+&(a_4-\frac{1}{2}a_{10})\left(\mathcal{M}_{ann}^{i(a,LL)}(Q_d)
+\mathcal{M}_{ann}^{i(b,LL)}(Q_d)\right) +(a_6-\frac{1}{2}a_8)
\mathcal{M}_{ann}^{i(SP)}(Q_d)\Big\}.\label{e60}
\end{eqnarray}
It is seen that the $O_{7\gamma}$
contributions to the $B^+$ and $B^0$ meson decays are identical.

The explicit expressions for some functions appearing in the above factorization
formulas are presented below. We adopt the model
\begin{eqnarray}
\phi_B(x,b)=N_Bx^2(1-x)^2\text{exp}\left[-\frac{1}{2}(\frac{xm_B}{\omega_B})^2-\frac{1}{2}(\omega_Bb)^2\right],
\end{eqnarray}
for the $B$ meson DA,
where $b$ is the impact parameter conjugate to the spectator transverse
momentum $k_T$, $N_B$ is the normalization constant, and the shape parameter
$\omega_B=0.40\pm0.04$ GeV has been determined through the study of the
$B$ meson transition form factors~\cite{Keum:2000ph,Keum:2000wi}.

The hard scales are chosen as
\begin{eqnarray}
t_{a}&=&{\rm max}\{ \sqrt{z}m_B, 1/b_1, 1/b_2 \}, \nonumber \\
t_{a}'&=&{\rm max}\{\sqrt{|x_1-\eta|}m_B, \sqrt{zx_1} m_B, 1/b_1, 1/b_2 \},\nonumber \\
t_{b}&=&{\rm max}\{\sqrt{1-z}m_B, \sqrt{zx_1} m_B, 1/b_1, 1/b_2 \},\nonumber \\
t_{b}'&=&{\rm max}\{\sqrt{1+x_1-\eta}m_B, \sqrt{zx_1} m_B, 1/b_1, 1/b_2 \},\nonumber \\
t_{c}&=&{\rm max}\{\sqrt{z(1-\eta)}m_B, \sqrt{|(x_1-z)(1-\eta)|} m_B, 1/b_1, 1/b_2 \},\nonumber \\
t_{c}'&=&{\rm max}\{\sqrt{x_1(1-\eta)}m_B, \sqrt{|(x_1-z)(1-\eta)|} m_B, 1/b_1, 1/b_2 \},\nonumber \\
t_{d}&=&{\rm max}\{ \sqrt{x_1z}m_B, \sqrt{|B^2|}, 1/b_1 \}, \nonumber \\
t_{e}&=&{\rm max}\{ \sqrt{1-\eta+x_1}m_B,  1/b_1 \},
\quad t_{e}'={\rm max}\{ \sqrt{x_1(1-\eta)}m_B,  1/b_1 \}, \nonumber \\
t_{f}&=&{\rm max}\{ \sqrt{1-z}m_B,  1/b_2 \}, \quad t_{f}'={\rm max}\{ \sqrt{z(1-\eta)}m_B,  1/b_2 \}.
\end{eqnarray}
The hard functions are written as
\begin{eqnarray}
h_{a}(x_1,z,b_1,b_2)&=&K_0( \sqrt{zx_1}m_Bb_1)[\theta(b_1-b_2)
K_0(\sqrt{z} m_Bb_1)I_0(\sqrt{z} m_Bb_2)+(b_1\leftrightarrow b_2)]S_t(z),\nonumber \\
h_{a}'(x_1,z,b_1,b_2)&=&K_0( \sqrt{zx_1}m_Bb_2)S_t(x_1)\nonumber \\
&&\times\{\begin{array}{ll} \frac{i\pi}{2}[\theta(b_2-b_1)
H_0^{(1)}(\sqrt{|x_1-\eta|}m_Bb_2)J_0(\sqrt{|x_1-\eta|}m_Bb_1)
+(b_2\leftrightarrow b_1)], &x_1<\eta,\nonumber \\
\theta(b_2-b_1)K_0(\sqrt{x_1-\eta}m_Bb_2)I_0(\sqrt{x_1-\eta}m_Bb_1)
+(b_2\leftrightarrow b_1), &x_1>\eta, \end{array} \nonumber \\
h_{b}(x_1,z,b_1,b_2)&=&K_0(\sqrt{x_1z} m_Bb_1)\frac{i\pi}{2}
[\theta(b_1-b_2)H_0^{(1)}(\sqrt{1-z} m_Bb_1)J_0(\sqrt{1-z} m_Bb_2)
+(b_2\leftrightarrow b_1)]S_t(z),\nonumber \\
h_{b}'(x_1,z,b_1,b_2)&=&K_0(\sqrt{x_1z} m_Bb_2)S_t(x_1)[\theta(b_2-b_1)
K_0(\sqrt{1+x_1-\eta} m_Bb_2)I_0(\sqrt{1+x_1-\eta} m_Bb_1)\nonumber \\
&&+(b_2\leftrightarrow b_1)],\nonumber \\
h_{c}(x_1,z,b_1,b_2)&=&\frac{i\pi}{2}[\theta(b_1-b_2)
H_0^{(1)}(\sqrt{z(1-\eta)} m_B b_1)J_0(\sqrt{z(1-\eta)} m_Bb_2)
+(b_2\leftrightarrow b_1)]S_t(z)\nonumber \\
&&\times\{\begin{array}{ll} \frac{i\pi}{2}
H_0^{(1)}(\sqrt{|(x_1-z)(1-\eta)|}m_Bb_1), &x_1<z,\nonumber \\
K_0(\sqrt{(x_1-z)(1-\eta)}m_Bb_1), &x_1>z, \end{array} \nonumber \\
h_{c}'(x_1,z,b_1,b_2)&=&[\theta(b_2-b_1)
K_0(\sqrt{x_1(1-\eta)} m_B b_2)I_0(\sqrt{x_1(1-\eta)} m_Bb_1)
+(b_2\leftrightarrow b_1)]S_t(x_1)\nonumber \\
&&\times\{\begin{array}{ll} \frac{i\pi}{2}
H_0^{(1)}(\sqrt{|(x_1-z)(1-\eta)|}m_Bb_2), &x_1<\eta,\nonumber\\
K_0(\sqrt{(x_1-z)(1-\eta)}m_Bb_2), &x_1>z. \end{array}
\end{eqnarray}
The Sudakov factor $S_t(x)$ from the threshold resummation
follows the parametrization in~\cite{Li:2002mi}
\begin{eqnarray}
S_t(x)=\frac{2^{1+2a}\Gamma(3/2+a)}{\sqrt{\pi}\Gamma(1+a)}[x(1-x)]^a,
\end{eqnarray}
with the parameter $a=0.4$.
The evolution factors are given by
\begin{eqnarray}
E_{e}(t)&=&\alpha_s(t){\rm exp}[-S_B(t)-S_2(t)], \nonumber \\
E_a(t)&=&S_t(x_1){\rm exp}[-S_B(t)], \quad E_a'(t)=S_t(z){\rm exp}[-S_2(t)],
\end{eqnarray}
with the Sudakov exponents
\begin{eqnarray}
S_B(t)&=&s(x_1\frac{m_B}{\sqrt{2}},b_1)+\frac{5}{3}\int_{1/b_1}^{t}
\frac{d\bar{\mu}}{\bar{\mu}}\gamma_q(\alpha_s(\bar{\mu})),\nonumber \\
S_2(t)&=&s(z\frac{m_B}{\sqrt{2}},b_2)+s((1-z)\frac{m_B}{\sqrt{2}},b_2)
+2\int_{1/b_2}^{t}\frac{d\bar{\mu}}{\bar{\mu}}\gamma_q(\alpha_s(\bar{\mu})),
\end{eqnarray}
$\gamma_q=-\alpha_s/\pi$ being the quark anomalous dimension.
The function $s(Q,b)$ is expressed as
\begin{eqnarray}
s(Q,b)&=&\frac{A^{(1)}}{2\beta_1}\hat{q} \text{ln}
\big(\frac{\hat{q}}{\hat{b}}\big)-\frac{A^{(1)}}{2\beta_1}(\hat{q}-\hat{b})
+\frac{A^{(2)}}{4\beta_1^2}\hat{q} \text{ln} \big(\frac{\hat{q}}{\hat{b}}-1\big)\nonumber \\
&-&\big[\frac{A^{(2)}}{4\beta_1^2}-\frac{A^{(1)}}{4\beta_1}
\text{ln} \big(\frac{e^{2\gamma_E-1}}{2}\big)\big] \text{ln}
 \big(\frac{\hat{q}}{\hat{b}}\big),
\end{eqnarray}
where
\begin{eqnarray}
\hat{q}&=& \text{ln} \frac{Q}{\sqrt{2}\Lambda_{\rm QCD}},
\quad \hat{b}= \text{ln} \frac{1}{b\Lambda_{\rm QCD}}, \nonumber \\
\beta_1&=&\frac{33-2n_f}{12}, \quad \beta_2=\frac{153-19n_f}{24}, \nonumber \\
A^{(1)}&=&\frac{4}{3}, \quad A^{(2)}=\frac{67}{9}-\frac{\pi^2}{3}
-\frac{10}{27}n_f+\frac{8}{3}\beta_1 \text{ln} \big(\frac{1}{2}e^{\gamma_E}\big),
\end{eqnarray}
$n_f$ being the number of the quark flavor, and $\gamma_E$ the Eular constant.

\end{document}